\documentclass{aa}  
\usepackage{graphicx}
\usepackage{txfonts}
\usepackage{natbib} \bibpunct{(}{)}{;}{a}{}{,} 
\usepackage{hyperref}
\begin{document} 
   \title{The diffuse molecular component in the nuclear bulge of the Milky Way \thanks{This paper makes use of the following ALMA data: ADS/JAO.ALMA\#2012.1.00119.S. ALMA is a partnership of ESO (representing its member states), NSF (USA) and NINS (Japan), together with NRC (Canada), NSC and ASIAA (Taiwan), and KASI (Republic of Korea), in cooperation with the Republic of Chile. The Joint ALMA Observatory is operated by ESO, AUI/NRAO and NAOJ.}}
   \author{D. Riquelme \inst{1} \thanks{This work was partially carried out during several visits of the principal author to the Astronomy Department of the Universidad de Chile.}
   \and L. Bronfman \inst{2}
   \and R. Mauersberger \inst{1,2}
   \and R. Finger \inst{2}
   \and C. Henkel \inst{1,3}
   \and T.L. Wilson \inst{4}
   \and P. Cort\'es-Zuleta \inst{2}}
   \institute{Max-Planck-Institut f\"ur Radioastronomie, Auf dem H\"ugel 69,
    53121 Bonn, Germany\\
    \email{riquelme@mpifr-bonn.mpg.de}
    \and Departamento de Astronom\'{i}a, Universidad de Chile, Casilla 36-D, Santiago, Chile
    \and Astronomy Department King Abdulaziz University, P.O. Box 80203, Jeddah 21589, Saudi Arabia
    \and NSF Division of Astronomical Sciences. Suite 1053, 4201 Wilson Blvd. Arlington VA 22230, USA}
    \date{Received; accepted}

 
  \abstract
   {The bulk of the Molecular gas in the Central Molecular Zone (CMZ) of the Galactic center region shows warm kinetic temperatures, ranging from $>20$ K in the coldest and densest regions (n$\sim 10^{4-5}$ cm$^{-3}$) up to more than 100 K for densities of about n$\sim 10^3$ cm$^{-3}$.
   Recently, a more diffuse, hotter ($n \sim 100$ cm$^{-3}$, $T\sim 250$ K)  gas component was discovered through absorption  observations of H$_3^+$. This component may be widespread in the Galactic center, and low density gas detectable in absorption may be present even outside the CMZ along sightlines crossing the extended bulge of the Galaxy.}  
   {We aim to observe and characterize diffuse and low density gas using observations of 3-mm molecular transitions seen in absorption.}
   {Using the Atacama Large (sub)Millimeter Array (ALMA) we observed the absorption against the quasar J1744-312, which is located towards the Galactic bulge region at $(l,b)=(-2{^\circ}.13,-1{^\circ}.0)$, but outside the main molecular complexes.}
   {ALMA observations in absorption against the J1744-312 quasar reveal a rich and complex chemistry in low density molecular and presumably diffuse clouds. We detected three velocity components, at $\sim $ 0, $-153$, and $-192$ km s$^{-1}$. The component at $\sim 0$ km s$^{-1}$ could represent gas in the Galactic disk while the velocity components at $-153$, and $-192$ km s$^{-1}$ likely originate from the Galactic bulge. We detected 12 molecules in the survey, but only 7 in the Galactic bulge gas.}
   {}


   \maketitle
%

\section{Introduction}
Until recently, our view of the interstellar medium (ISM) of the Central Molecular Zone (CMZ, i.e. the innermost 500 pc) of the Milky Way was that it consists of basically three phases: an ionized and low density gas component ($T\sim  10^{4-6}$ K, n$\sim 10$ cm$^{-3}$)  responsible for the well-studied fine structure and radio recombination line emission; an ultra high temperature X-ray-emitting plasma which covers a large fraction of the CMZ, and dense and warm giant molecular clouds, which amount to about $10^7$ M$_{\odot}$,  and contain $\sim 80 \%$ of the dense gas \citep{Morris_Serabyn_1996,Lazio_Cordes_1998, Geballe_et_al_2011, Ferriere_et_al_2007}. This molecular phase, shows warm gas with kinetic temperatures 
ranging from $\sim 20$ K in the coldest and densest regions (n$\sim 10^{4-5}$ cm$^{-3}$) up to more than 100 K when densities decrease to n$\sim 10^3$ cm$^{-3}$ \citep{Huttemeister_et_al_1993}. Recently, using H$_2$CO observations, \citet{Ginsburg_et_al_2016} showed that warm (50-120 K) and dense gas (n$\sim 10^{4-5}$ cm$^{-3}$) pervades the CMZ. However, detections of H$_3^+$ absorption \citep{Goto_et_al_2002} led to the discovery of a vast amount of high-temperature ($T\sim 250$ K) and low density ($n \sim 100$ cm$^{-3}$) gas with a large velocity dispersion in the CMZ  \citep{Oka_et_al_2005, Goto_et_al_2008,Geballe_Oka_2010, lePetit_et_al_2016}. This new component presumably represents a diffuse molecular medium which is widespread in the CMZ, distinct from the three ISM components already known. Although the absorbing gas seems to be ubiquitous toward the  Galactic center region, little is known about its volume filling factor, mass, the detailed excitation conditions ($n({\rm H}_2$), $T_{\rm kin}$), chemical composition and the extension of it. This diffuse molecular component has been detected very recently by \citet{Gerin_Liszt_2017} in three species, namely HCO$^+$, HCN and HNC.

The new component had largely escaped detection in large scale surveys in the cm and mm-range, because molecules, in order to be detectable in emission, require a) a sufficient abundance in the upper energy levels and b) a dipole moment. CO is often used as a tracer of H$_2$ since it does have a dipole moment, even a sufficiently small one so that its rotational levels can be populated by collisions with H$_2$ when densities are less than 1000 cm$^{-3}$. But what if densities are smaller than a few 100 cm$^{-3}$ and the excitation temperature of even the lower CO rotational transitions comes close to the cosmic microwave background? In that case, emission from CO (and other molecules) would become completely invisible due to the lack of excitation above the cosmic background emission. 

Fortunately, one can detect such low excitation gas, namely via its absorption lines. Toward the disk of the Milky Way, one can e.g. trace diffuse/translucent gas by looking for absorption lines toward quasars which happen to be within 20 degrees of the Galactic plane \citep{Liszt_Lucas_2001}. A surprisingly complex polyatomic chemistry exists in such diffuse clouds, allowing for detections of species such as C$_2$H, C$_3$H$_2$, H$_2$CO, and NH$_3$, which have relative abundances that are strikingly similar to those of the Taurus molecular cloud 1 \citep[TMC-1, ][]{Liszt_et_al_2008}. In extragalactic studies, absorption against the quasar PKS1830-211 reveals a rich chemistry in the disk of a galaxy at z =0.89 \citep{Muller_et_al_2011, Muller_et_al_2016,Menten_et_al_1999, Menten_et_al_2008}. 
\citet{Muller_et_al_2011} measured many absorption lines finding excitation temperatures of almost all species involved around 5.1 K, i.e. the CMB temperature at that distance. These lines would therefore be impossible to detect in emission. \citet{Henkel_et_al_2008} showed from NH$_3$ absorption lines toward the same quasar that despite the low excitation, the kinetic temperature of the gas ranges from $T=80$ up to $600$ K, and may therefore be similar in excitation conditions to the component claimed by \citet{Geballe_Oka_2010} for our Galactic center region. Also toward the CMZ, 
ammonia absorption against Sgr\,B2 reveals low excitation gas with a kinetic temperature exceeding 1300 K \citep{Wilson_et_al_2006}. \citet{Wilson_et_al_2006} assumed that this  hot gas is physically associated with the active region Sgr\,B2. Following the claim by \citet{Geballe_Oka_2010}, it may well be that what Wilson et al. observed is indeed hot, low excitation, and most important: diffuse and widespread inter-cloud gas in the Galactic center region. 

Here we intend to search for the presence of the low density gas by looking for absorption toward a source along a sightline beyond the CMZ, crossing the bulge of the Galaxy and not being associated with a giant molecular cloud. There are still many unknowns about this new component such as excitation condition and molecular composition. Therefore we have carried out a pilot study, to detect molecular absorption against the quasar J1744-312 located behind the bulge of the Galaxy in order to confirm and characterize the diffuse molecular component in the Galactic center region proposed by Geballe \& Oka.

\subsection{The quasar J1744-312}\label{thequasar}
\begin{figure*}
\centering
\includegraphics[width=0.22\hsize, angle=90]{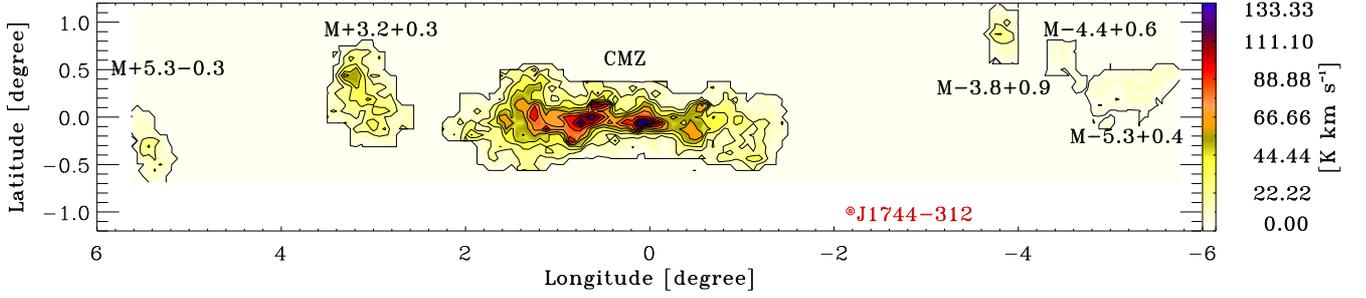}
\caption{Overview of the Galactic center region in the HCO$^+$ (1-0) line from \citet{Riquelme_et_al_2010b}. The position of the quasar J1744-312 is indicated.}
\label{overview}
\end{figure*}

We have chosen the quasar J1744-312 as the background source to look for the absorption components in the Galactic center (GC) region.  J1744-312 is at $(l, b) = (-2{^\circ}.13, -1{^\circ}.00)$ i.e., behind the Galactic bulge \footnote{Throughout this work, we refer as the Galactic center region to the central kpc of the Galaxy, which is shown in Fig. \ref{overview} and which is related to the bulge of the Galaxy}. Fig. \ref{overview} shows the position of the quasar in the GC region. This quasar is compact, since it is at best marginally resolved with a 4.5x1.1 mas beam at 24 GHz \citep[B1741-312 in ][]{Charlot_et_al_2010}, and it is expected that the quasar is even more compact at mm-wavelengths. In addition interstellar scintillation, which can result in a broadening of the source size, is negligible in our case since the quasar is far from the CMZ and scintillation becomes less visible at mm-wavelengths. The quasar has a flux density of 0.571 Jy at 103 GHz and 0.563 Jy at 91.5 GHz in the ALMA-Band 3  calibrator source catalog (see the ALMA Calibrator Source Catalogue  \footnote{\url{https://almascience.eso.org/sc/}}); it is therefore an adequate source for absorption measurements of gas in the Galactic bulge. Unlike the absorption detected by \citet{Geballe_Oka_2010}, this quasar is not behind one of the prominent molecular complexes. This may serve to avoid confusion with absorption by denser gas components. 

\section{Observations}
The observations were carried out during the ALMA Early Science Cycle 1 (ALMA project 2012.1.00119.S) using 29 12-m antennas during 5 observing periods (May 02; Jun 29; Jul 01, 02, 03, and 21, 2014). The coordinates for the quasar used for the observations were $\alpha$(J2000): $17^{\rm h}44^{\rm m}23.5782^{\rm s}$, $\delta$(J2000): $-31\deg16'36''.287$.
We performed a spectral survey of the ALMA band 3, covering a frequency range from 84.114 to 114.616 GHz using five spectral setups, each of them covering 4x1.875 GHz bandwidth with a channel separation of $2.584$ MHz corresponding to $\sim 10$ km s$^{-1}$. The primary beam full-width at half-maximum (FWHM) size ranges from 73.5 arcsec at 84.114 GHz to 54 arcsec at 114.616 GHz. Table \ref{table:observation} shows the details for the observations and the sensitivity reached in each setup. 
The calibration was performed by the ALMA team at the North America Arc node. 
     
\begin{table*}
\caption{Observational setups}             
\label{table:observation}      
\centering                          
\begin{tabular}{ccccccccc}        
\hline\hline                 
setup & spectral & Frequency range &Central Frequency & rms & Velocity &Primary  & Syntetized  & Observation date\\    
      & window   &                &   	           &     &  resolution &beam    &  beam     &\\
      &          &   GHz          &   GHz	   & mJy & km\,s$^{-1}$& arcsec       &        arcsec   &\\ 
\hline                        
1 & 0 & 91.483-93.355   & 92.416 & 2.56 & 8.42 & 66.87 & 1.55x1.36  &  2014-06-29\\    
  & 1 & 93.316-95.188   & 94.264 & 2.37 & 8.42 & 65.56 & 1.45x1.34  &   \\    
  & 2 & 103.627-105.498 &104.563 & 2.59 & 8.42 & 59.10 & 1.34x1.24  &   \\    
  & 3 & 105.483-107.354 &106.412 & 2.56 & 8.42 & 58.07 & 1.35x1.23  &   \\    
2 & 0 & 95.158-97.029	& 96.090 & 3.19 & 8.13 & 64.31 & 1.91x1.35  &  2014-07-01 \\    
  & 1 & 96.991-98.860	& 97.934 & 3.20 & 8.13 & 63.10 & 1.87x1.36  &   \\    
  & 2 & 107.303-109.173	&108.240 & 3.77 & 8.13 & 57.09 & 1.74x1.31  &   \\    
  & 3 & 109.158-111.030	&110.090 & 3.83 & 8.13 & 56.14 & 1.70x1.28  &   \\    
3 & 0 &  87.810-89.681	& 88.742 & 2.94 & 8.73 & 69.64 & 1.80x1.49  & 2014-07-03 \\
  & 1 &  89.643-91.513	& 90.590 & 3.19 & 8.73 & 68.22 & 1.75x1.50  &  \\                
  & 2 &  99.954-101.825	&100.890 & 3.05 & 8.73 & 61.25 & 1.63x1.43  &  \\      
  & 3 & 101.810-103.680 &102.740 & 3.24 & 8.73 & 60.15 & 1.61x1.41  &  \\     
4 & 0 & 110.925-112.798 &111.861 & 2.99 & 8.83 & 55.25 & 1.55x1.26  & 2014-07-02 \\  
  & 1 & 112.747-114.616 &113.695 & 4.01 & 8.83 & 54.36 & 1.53x1.23  &  \\
  & 2 &  98.789-100.660 & 99.714 & 2.46 & 8.83 & 61.98 & 1.76x1.36  &  \\
  & 3 & 100.611-102.480 &101.547 & 2.32 & 8.83 & 60.86 & 1.63x1.33  &  \\
5 & 0 &  84.114-85.994  & 85.067 & 1.35 & 12.04& 72.65 & 1.70x.140  &  2014-05-02, 2014-07-21\\
  & 1 &  85.970-87.850  & 86.916 & 1.47 & 11.79& 71.10 & 1.65x1.41  &  \\
  & 2 &  96.280-98.160  & 97.216 & 1.53 & 10.54& 63.57 & 1.56x1.33  &  \\  
  & 3 &  98.113-99.993  & 99.064 & 1.56 & 10.34& 62.38 & 1.51x1.33  &  \\\hline
\hline        
\end{tabular}    
\end{table*}    

\section{Results and analysis}

The analysis of the data was done with the GREG and CLASS packages of the GILDAS software\footnote{\url{http://www.iram.fr/IRAMFR/GILDAS}}. Our data consist of absorption lines and continuum levels toward the quasar. We extract a single spectrum toward the center of the quasar. The continuum level ranges from 0.55 to 0.70 Jy  with small variations not only along the band ($<5$\%), but also between the different observing periods ($<14$\%), which suggests intrinsic intensity variations of J1744-312. Then, to compute the optical depth, we estimate the line-to-continuum ratio for each detected line (see Section \ref{columndensity}).

\subsection{Line identification and line fit} \label{fitting}
\begin{figure*}
\centering
\includegraphics[width=\hsize]{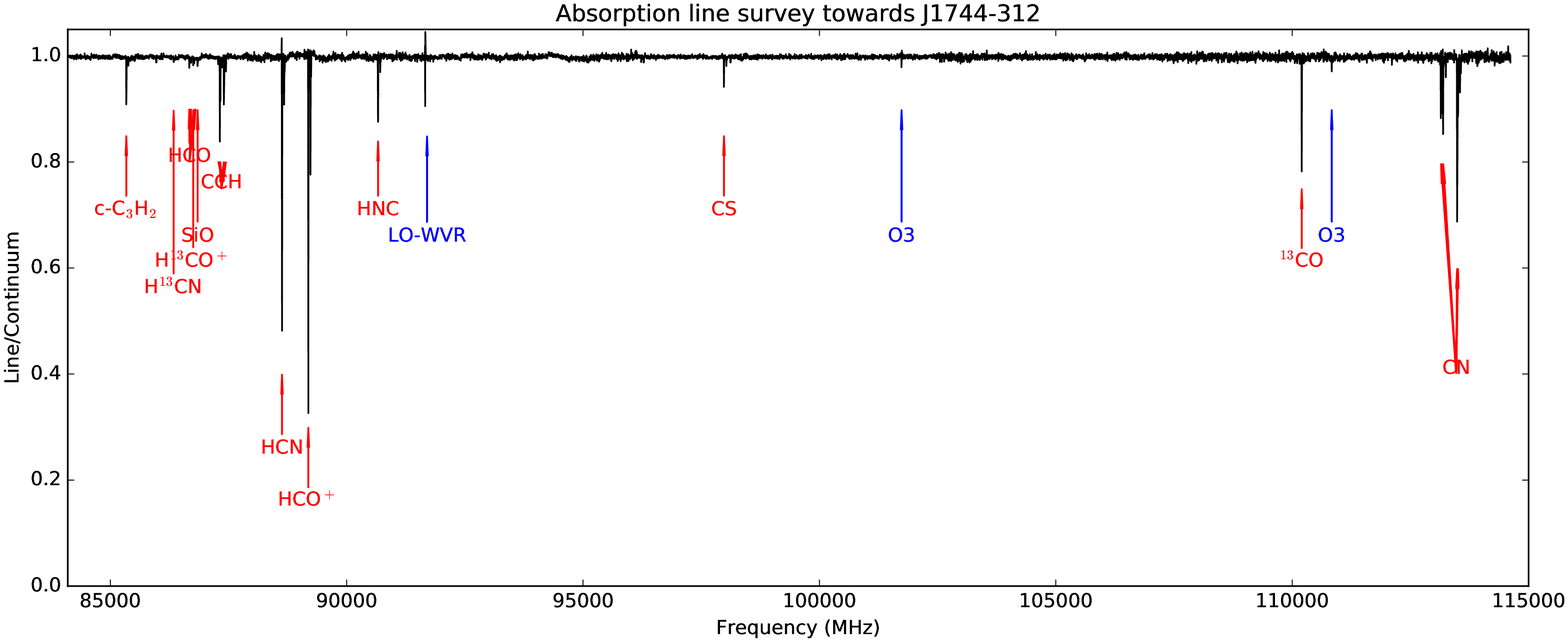}
\caption{3-mm survey toward J1744-312. The signal has been normalized to the continuum level. In red all detected interstellar molecular transitions are displayed. In blue we show the atmospheric O$_3$ lines and interference by the local oscillators of the water vapor radiometers  (see Sect. \ref{fitting}).}
\label{survey}
\end{figure*}

Fig. \ref{survey} shows the complete spectrum normalized to the detected continuum level. The line identification was done using the Cologne Database for Molecular Spectroscopy (CDMS) catalog \citep{Muller_et_al_2005, Muller_et_al_2001}, and the Jet Propulsion Laboratory Molecular Spectroscopy (JPL) catalog \citep{Pickett_et_al_1998}. Twelve species were detected, including three molecules with  hyperfine structure (HFS), namely HCO, CN, and CCH\footnote{HCN also have HFS but it is not resolved because of the poor spectral resolution}; one instrumental artifact at 91.7 GHz, and two atmospheric transitions at 101.74 and  110.836 GHz. Table \ref{table:1} shows the spectral  parameters of all detected interstellar molecular transitions. 

The features close to 91.65 GHz, which are seen in emission and absorption, can be ascribed to interference by the local oscillators (LO) of the water vapor radiometers (WVR) each of the ALMA antennas is equipped with \citep{Nikolic_et_al_2013}. The signal of the LO, if insufficiently shielded, may be picked up by one or several neighboring ALMA antennas. Even though each  LO in the different WVRs is tuned to a slightly different frequency around 91.7 GHz in order to avoid a correlated product, this measure is apparently not sufficient to prevent an interference free signal at a frequency close to 91.7 GHz. The features at 110.836  and 101.74 GHz correspond to atmospheric O$_3$ \citep{Shimabukuro_Wilson1973,Caton_et_al_1968}. 

In order to characterize the absorption profile with a higher signal-to-noise level, we averaged spectra of the four strongest transitions that do not have HFS. This stacked spectrum clearly shows three velocity components at $\sim $ 0, $-153$, and $-192$ km s$^{-1}$ (see Fig. \ref{stack}). 
The component at $\sim 0$ km s$^{-1}$ could correspond to gas in the Galactic disk, and we argue in Section \ref{location} that the velocity components at $-153$, and $-192$ km s$^{-1}$ are from the bulge of the Galaxy.  The absorption profiles were fitted with Gaussian functions  (Table \ref{table:2}). The intensity peak is negative because the fits are performed on the continuum subtracted spectra. The Appendix (Figs. \ref{fit} and \ref{FigVibStab}) shows Gaussian fits for each detected species. Because of the low velocity resolution of the observations, the fits did not always converge, and in those cases we fixed the linewidths to the value given by the Gaussian fit isolating each corresponding velocity component (Table \ref{table:2}). 
We clearly detected 4 HFS components of HCO, 6 HFS components of CCH, and 7 of the 9 HFS components of CN. These lines were modeled using the HFS method of CLASS assuming optically thin emission, and the same excitation temperature and linewidth for all components of the multiplet.  The relative intensities of the HFS of these molecules are shown in Table \ref{table:hfs}.  Only the velocity component at 0 km s$^{-1}$ was detected in HCO. Two velocity components were detected in CN and CCH. The component at $\sim 0$ km s$^{-1}$ is shown in red in Figs. \ref{fit} and \ref{FigVibStab}, and in blue the velocity component at $\sim -153$ km s$^{-1}$. The main HFS component is used to determine the total optical depth for HCO and CCH because they are not blended with any other species. CN shows a more complex spectrum, with three HFS features blended close to the main component. Here the CN line at 113.191 GHz, which is not blended with any other HFS feature (see Table \ref{table:hfs} and Fig. \ref{FigVibStab}), was used to determine the column densities (see Section \ref{columndensity}).

We compared the observed with the predicted line intensity (S$_l$, flux density peak) ratios of different HFS components in the optically thin case under Local Thermodynamic Equilibrium (LTE) conditions (Table \ref{table:2}). For CCH, the theoretical line ratio between the main line at 87.31689 GHz and the satellites at 87.32858 and 87.40198 GHz (which have the same relative intensity of 0.208, see Table \ref{table:2}) is 2.0, and the measured line intensity ratios (Gaussian peaks) are 1.9 and 2.0, respectively. For HCO, the theoretical line ratio between the main line at 86.67076 GHz and the satellites at 86.70836 and 86.77746 GHz (both with the same relative intensity) is 1.7, while the measured line intensity ratios are 2.0 and 1.8, respectively. For CN, the main line is blended with other satellites; then, to compute the ratio, we used the line at 113.19127 GHz with respect to the lines at 113.17049 GHz  and  113.14415 GHz, which have a theoretical line ratio of 1.3 and a measured line intensity ratio of 1.3 and 1.2, respectively. These results indicate that we can assume that CCH, CN and HCO are optically thin following LTE conditions, and that our estimation of the column densities is reliable under this approach. Therefore, the determination of the optical depth and then, the column density, was done using only the main HFS component for HCO and CCH, and the satellite at 113.191 GHz for CN, multiplying the obtained values by the proper factor (assumed: local        thermodynamical equilibrium, optically thin  absorption; see Sect. 3.2) to also account for the other HFS features. The corresponding molecular parameters are shown in Table \ref{table:1}.

In the spectra of some of the stronger lines (HCN, HCO$^+$ and CN) there is a one channel wide emission adjacent to the absorption. This is most probably an artifact due to the correlation of steep spectral features, known as Gibbs ringing \citep{Rupen_1999}.

\subsection{Physical properties} \label{columndensity}

The derivation of the column density from absorption lines, from single dish observations, has been discussed by \citet{Linke_et_al_1981} and \citet{Nyman_1984}.  
Here we use ALMA,  the most powerful (sub-)mm synthesis instrument, to take point source absorption spectra toward a target having a $\sim 0.6$ flux density and milliarcsec size, to thus obtain accurate optical depth measurements. 
More recent derivations of column density from absorption lines data, using ALMA observations, can be found in \citet{Ando_et_al_2016} and \citet{Muller_et_al_2014}.

Realistically assuming that the difference between excitation temperature and that of the cosmic microwave background is small in relation to the brightness temperature of the background quasar (actually, not only the difference but also $T_{ex}$ and $T_{bg}$ themselves are small with respect to $T_{\rm quasar}$) and that the absorption covers the entire background source, the optical depth ($\tau$) is related to the line/continuum flux density ratio by

\begin{equation}\label{tau}
      \tau = -ln\bigg(1+\frac{S_{\rm l}}{S_{\rm c}}\bigg) \,,
\end{equation}

where $S_{\rm l}$ is the flux density peak of the absorption line measured from the continuum level, and $S_{\rm c}$ is the flux density of the background continuum source. If all the molecular lines are spectroscopically resolved, then the $\tau$ estimated in this way is a good determination for the optical depth at the line peak.

However, even in the case of low spectral resolution, the line equivalent width (velocity integrated opacity in the optically thin case) is preserved and the total column density can be obtained with 

\begin{equation}\label{N}
      N=\frac{8\pi\nu^3}{c^3A_{ul}g_u}\frac{Q(T_{\rm ex}) \exp{\frac{E_l}{k_B T_{\rm ex}}}}{[1-\exp(\frac{-h\nu}{k_BT_{\rm ex}})]}\int \tau dV \, [cm^{-2}] .
\end{equation}
$k$ is the Boltzmann constant, $\nu$ the frequency of the transition, $h$ the Planck constant, 
$Q(T)$ the partition function at the assumed excitation temperature, $g_u$ the upper state degeneracy, $A_{ij}$ the Einstein coefficient, and $E_l$ the energy of the lower level.

Table \ref{table:2} shows the results for all the absorption features detected in the survey. As mentioned in section \ref{fitting}, Gaussian fits were used; then, in Eq. \ref{N}, $\int \tau dV = \sqrt{\frac{\pi}{4 ln(2)}}\tau_{\rm peak} \Delta V$, where $\tau_{\rm peak}$ is the $\tau$ at the peak of the absorption fitted by the Gaussian. Since the HFS line ratios are compatible with the theoretical LTE optically thin values (see section  \ref{fitting}), opacities (e.g. due to spectrally unresolved narrow features) can not be much larger than those given in Table \ref{table:2}. This in turn makes clear that the background source coverage factor must indeed be close to unity.

\begin{table*}
\caption{Line parameters and fitting results for the detected species.}             
\label{table:2}      
\centering                          
\begin{tabular}{c c c c c c c c c}        
\hline\hline                 
Molecule & Rest. Frequency & $V_0$ (LSR) & $\Delta V$ & $S_{\rm l}$ & $S_{\rm c}$& rms & $\tau$ & N \\    
	 & GHz &(km s$^{-1}$) & (km s$^{-1}$)&  Jy & Jy & mJy&      & (cm$^{-2}$)   \\
\hline  
c-C$_3$H$_2$&85.34 &-1.0 (0.2) & 18.6$^a$    & -0.06615 & 0.6195 & 1.244 & 0.113 (0.012) & $1.362 (0.400) \times 10^{13}$\\
	 & & -153.1 (2.2) & 15.0$^a$    & -0.00925 & 0.6195 & 1.244 & 0.015 (0.002) & $1.464 (0.512) \times 10^{12}$\\
	 & & -191.9 (2.1) & 12.0$^a$    & -0.00583 & 0.6195 & 1.244 & 0.009 (0.001) & $7.362 (3.261) \times 10^{11}$\\
H$^{13}$CN& 86.34& -2.0 (2.8) & 26.7 (6.3)  & -0.00535 & 0.6165 & 1.228 & 0.009 (0.001) & $4.860 (1.395) \times 10^{11}$\\
	 & & -&- &- & & &  &\\
	 & & -& -&- & & &  &\\
HCO$^b$  &86.71 &   -0.2 (0.8) & 19.7 (0.5)  & -0.01175$^b$ &0.6238 &1.244 & 0.019 (0.002) &$4.117 (4.220) \times 10^{12}$\\
	 & &- & & & & &  &\\
	 & &- & & & & &  &\\
H$^{13}$CO$^+$& 86.75 &0.2 (1.3)& 22.7 (3.8) & -0.0101   & 0.62429 & 1.095& 0.016 (0.002) &$4.538 (0.882)\times 10^{11}$\\
	 & & -&- & & & &  & \\
	 & & -&- & & & &  & \\
SiO      &86.85 &  2.4 (2.4)   & 22.7 (7.2)   & -0.00965 & 0.62401 &1.103 & 0.016 (0.002) & $1.283 (0.427) \times 10^{12}$\\
	 & & 		&              &          &         &      & & \\
	 & & 		&              &          &         &      & & \\
CCH$^b$  & 87.33&    -1.6 (0.8) & 18.499 (2.6) & -0.11309 & 0.62660 &1.260 & 0.199 (0.021) &$2.744 (0.480) \times 10^{14}$ \\
	 & &  -154.4 (1.5) & 26.680 (4.6) & -0.01248 & 0.62660 &1.260 & 0.020 (0.002)&$4.000 (0.795)\times 10^{13}$  \\
	 & &               &              &          &         &      &              & \\
HCN	 &88.63 &     0.5 (0.1) & 21.0 (0.1)   &-0.39488  & 0.65866 &2.953 & 0.915 (0.052)& $3.907 (0.224)\times 10^{13}$\\
	 & &  -154.2 (0.4) & 10.8 (1.1)   &-0.08517  & 0.65866 &2.953 & 0.138 (0.014)& $3.039 (0.438) \times 10^{12}$\\
	 & &  -193.3 (1.2) & 10.3 (2.0)   &-0.03140  & 0.65866 &2.953 & 0.049 (0.005)& $1.023 (0.222)\times 10^{12}$\\
HCO$^+$  & 89.19&     0.7 (8.7) & 20.0$^a$     &-0.47096  & 0.66217 &3.351 & 1.242 (0.058)& $2.936 (0.747) \times 10^{13}$\\
	 & &  -153.4 (8.7) & 10.6 (8.7)   &-0.18489  & 0.66217 &3.351 & 0.327 (0.029)& $4.114 (3.386) \times 10^{12}$ \\
         & &  -194.4 (8.7) & 10.0 (8.7)   &-0.04435  & 0.66217 &3.351 & 0.069 (0.007)& $8.170 (0.717) \times 10^{11}$\\
HNC 	 & 90.66&     1.5 (0.4) & 22.5 (0.6)   &-0.08730  & 0.66003 &2.827 & 0.142 (0.014) & $6.061 (0.641) \times 10^{12}$\\
	 & & -154.1 (0.5) &  8.7 (1.8)   &-0.03176  & 0.66003 &2.827 & 0.049 (0.005) & $8.160 (1.863)\times 10^{11}$\\
	 & & -            & -            &  -       &         &      &       &\\
CS       & 97.98&    3.9 (0.3) & 18.0$^a$     &-0.03808  & 0.64514 &1.353 & 0.061 (0.007) & $9.423 (2.817) \times 10^{12}$\\
	 & & -157.2 (0.9) & 10.0$^a$     &-0.01782  & 0.64514 &1.353 & 0.028 (0.003) & $2.410 (1.229) \times 10^{12}$\\
	 & & & & & & & & \\
$^{13}$CO& 110.20&   0.4 (0.1)  &  8.1 (0.7) & -0.26448   & 0.54710 & 3.717  & 0.660 (0.045) & $6.334 (0.694) \times 10^{15}$\\
	 & & & & & & & &\\
	 & & & & & & & &\\
CN$^b$   & 113.14&    0 (0.1)   & 14.7 $^a$& -0.11254$^c$ & 0.62611  &3.890 &  0.198 (0.019) & $1.547 (0.547) \times 10^{14}$\\
	 & & -156.0 (0.1) & 14.7$^a$ & -0.01906$^c$ & 0.62611  &3.890 &  0.031 (0.004) & $2.413 (0.890) \times 10^{13}$ \\
	 & &- &- & & & & & $<1.200\times 10^{13}$\\
\hline                                   
\end{tabular}
\tablefoot{V$_0$ is the central velocity (Local Standard of Rest, LSR) from a Gaussian fit, $\Delta$V is the full width at half maximum of the Gaussian fit, $S_{\rm l}$ corresponds to the flux density peak of the absorption line, $S_{\rm c}$ is the continuum flux density,  rms is the root-mean square noise level of the spectrum for a $\sim 10$ km s$^{-1}$ channel width,  $\tau$ is the optical depth,  $N$ is the column density. $^a$ Fixed parameter in the Gaussian fit. $^b$ Hyperfine structure. $^c$ Line intensity corresponds to the hyperfine component at 113.191 GHz. }
\end{table*}

\begin{table*}
\caption{Theoretical relative intensities of the detected hyperfine transitions assuming optically thin lines and local thermodynamical equilibrium. The intensities are normalized to unity}             
\label{table:hfs}      
\centering                          
\begin{tabular}{cccc}        
\hline\hline                 
Molecule & Transition &Res. Frequency & Relative Intensity\\    
	 & N'-N'', J'-J'', F'-F''  &[GHz]		&  \\
\hline                        

CN  & $1-0$, $1/2-1/2$, $1/2-1/2$ & 113.12337   &  0.012  \\    
    & $1-0$, $1/2-1/2$, $1/2-3/2$ & 113.14415   &  0.098  \\    
    & $1-0$, $1/2-1/2$, $3/2-1/2$ & 113.17049   &  0.096  \\    
    & $1-0$, $1/2-1/2$, $3/2-3/2$ & 113.19127   &  0.125  \\    
    & $1-0$, $3/2-1/2$, $3/2-1/2$ & 113.48812   &  0.125  \\    
    & $1-0$, $3/2-1/2$, $5/2-3/2$ & 113.49097   &  0.333  \\    
    & $1-0$, $3/2-1/2$, $1/2-1/2$ & 113.49964   &  0.099  \\    
    & $1-0$, $3/2-1/2$, $3/2-3/2$ & 113.50890   &  0.096  \\    
    & $1-0$, $3/2-1/2$, $1/2-3/2$ & 113.52043   &  0.012 \\\hline 
HCO & $1-0$, $3/2-1/2$, $2-1$ & 86.67076    & 0.420  \\                
    & $1-0$, $3/2-1/2$, $1-0$ & 86.70836    & 0.248  \\      
    & $1-0$, $1/2-1/2$, $1-1$ & 86.77746    & 0.248  \\     
    & $1-0$, $1/2-1/2$, $0-1$ & 86.80578    & 0.084  \\\hline     
CCH & $1-0$, $3/2-1/2$, $1-1$ & 87.28410    & 0.042  \\
    & $1-0$, $3/2-1/2$, $2-1$ & 87.31689    & 0.416  \\
    & $1-0$, $3/2-1/2$, $1-0$ & 87.32858    & 0.208  \\
    & $1-0$, $1/2-1/2$, $1-1$ & 87.40198    & 0.208  \\
    & $1-0$, $1/2-1/2$, $0-1$ & 87.40716    & 0.084  \\
    & $1-0$, $1/2-1/2$, $1-0$ & 87.44647    & 0.042  \\\hline
\hline        
\end{tabular}    
\end{table*}

\section{Discussion}\label{discussion}

\subsection{Association of the absorption features with the Galactic center region}\label{location}
\begin{figure}
\centering
\includegraphics[width=\hsize]{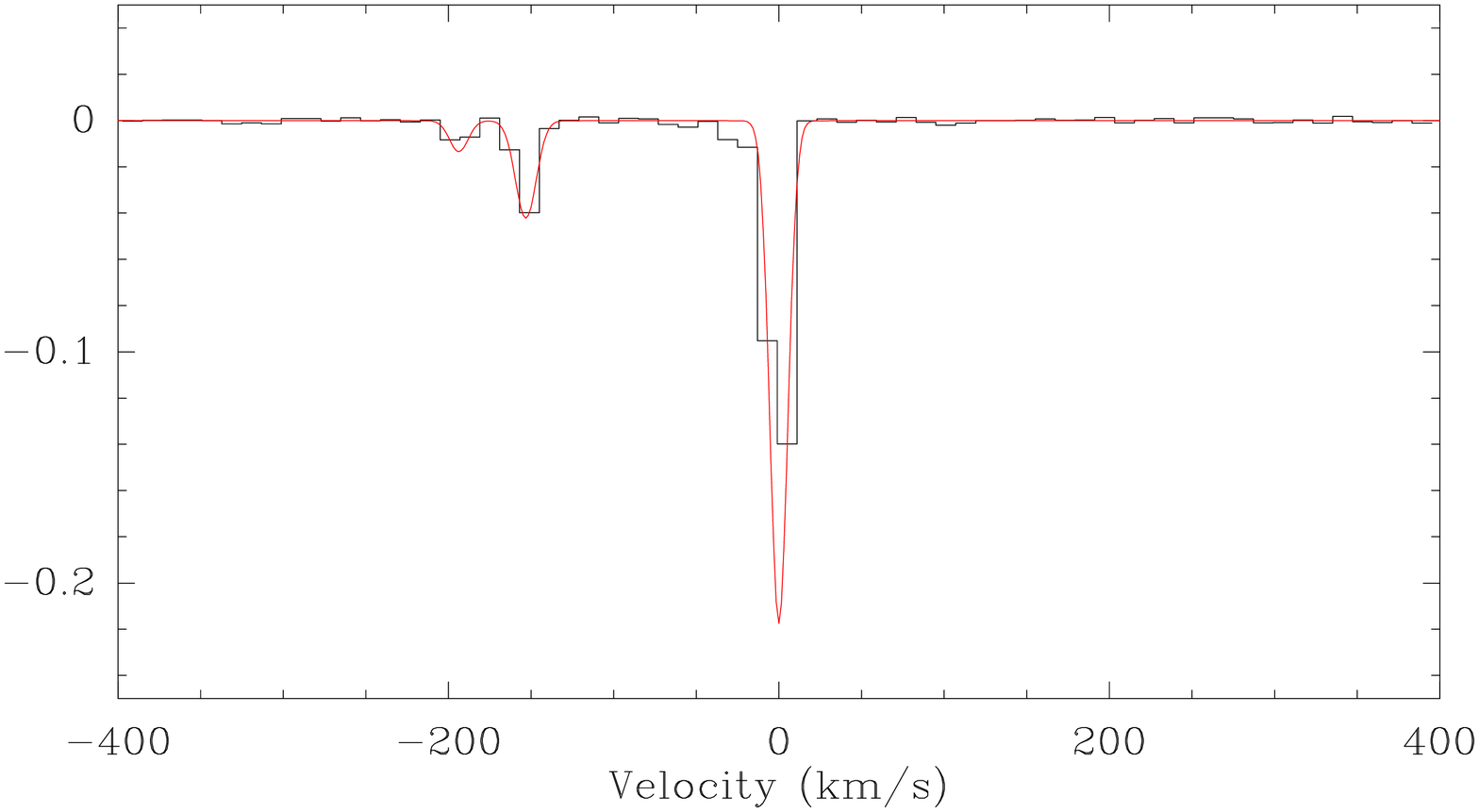}
\caption{Stacked spectrum of the four strongest transitions that show the three velocity components and do not have hyperfine structure (HCO$^+$, HNC, CS, c-C$_3$H$_2$). }
\label{stack}
\end{figure}

As stated in Section \ref{fitting}, there are three velocity components, at $\sim $ 0, $-153$, and $-192$ km s$^{-1}$ (Fig. \ref{stack}). 
The velocity component at 0 km s$^{-1}$ may be associated with gas along the line-of-sight across the Galactic disk as can be seen in the CO (1-0) maps from \citet{Bitran_et_al_1997} and the CO (2-1) emission line maps from \citet{Sawada_et_al_2001}. This velocity component is the only one detected also in the $^{13}$C isotopic substitutions of HCO$^+$ and HCN, translating into a $^{12}$C/$^{13}$C isotopic ratio of 65 (HCO$^+$/H$^{13}$CO$^+$) and 80 (HCN/H$^{13}$CN) assuming that both transitions are optically thin. Even though \citet{Riquelme_et_al_2010a} have found high $^{12}$C/$^{13}$C isotopic ratios in a few specific regions in the GC where gas may be infalling, the high isotopic values found in the 0 km s$^{-1}$ component are an indication that this gas represents disk gas. The isotope ratios match those found in the local interstellar medium (70) and  are much larger than the values of 20-25 usually found in the GC \citep{Wilson_1999, Milam_et_al_2005}. There were no $^{13}$C isotopic substitutions detected in the velocity components at $-153$, and $-192$ km s$^{-1}$. The lower limits (3$\sigma$) of HCO$^+$/H$^{13}$CO$^+$ and HCN/H$^{13}$CN in the $-153$ km s$^{-1}$ component are 42 and 16, respectively. For the $-192$ km s$^{-1}$ component, these are 11 and 7. 
There are no foreground features expected from, e.g., spiral arms at velocities of $-153$, and $-192$ km s$^{-1}$ \citep{Bania_1977, Dame_Thaddeus_2008}. 
There is no CO emission counterpart for the velocity components at $-153$ and $-192$ km s$^{-1}$ from previous observations at the position of the quasar \citep{Sawada_et_al_2001, Bitran_et_al_1997}. This is also noted by \citet{Gerin_Liszt_2017} who found HI emission around $-190$ km s$^{-1}$ and significantly weaker HI emission at $-150$ km s$^{-1}$ but CO emission only at $-22$ km s$^{-1}$ along the line-of-sight to the quasar. They only found CO emission from the bulge at $\sim -190$ km s$^{-1}$ in the vicinity of the quasar J1744-312 at lower Galactic latitudes ($b>-0{^\circ}\!.5$), and the velocity component at $-150$ km s$^{-1}$ only in a nearby CO feature.  This indicates that these absorption features correspond to low excitation gas that cannot be observed in emission towards the bulge of the Galaxy.

\subsection{Detected molecules}
From the 12 species observed in this study, 7 molecules were detected in the velocity component at $-153$ km s$^{-1}$, namely c-C$_3$H$_2$, CCH, HCN, HCO$^+$, HNC, CS, and CN. Out of those, only 3 were detected at $-192$ km s$^{-1}$ (c-C$_3$H$_2$, HCN and HCO$^+$). 

In previous absorption studies against quasars, \citet{Lucas_Liszt_2000} and \citet{Liszt_Lucas_2001} found that the molecules can be grouped into different chemical families: 1) the cyanogen- or CN-bearing molecules, including CN, HCN and HNC. The abundances of these molecules are tightly linearly correlated. 2) the HCO$^+$-OH family, which also shows a tight linear relation \citep{Liszt_Lucas_1996, Lucas_Liszt_1996}, and 3) the C$_n$H$_m$-family, with a looser but still good linear correlation. 

\begin{figure}
\centering
\vbox{
\includegraphics[width=0.99\hsize]{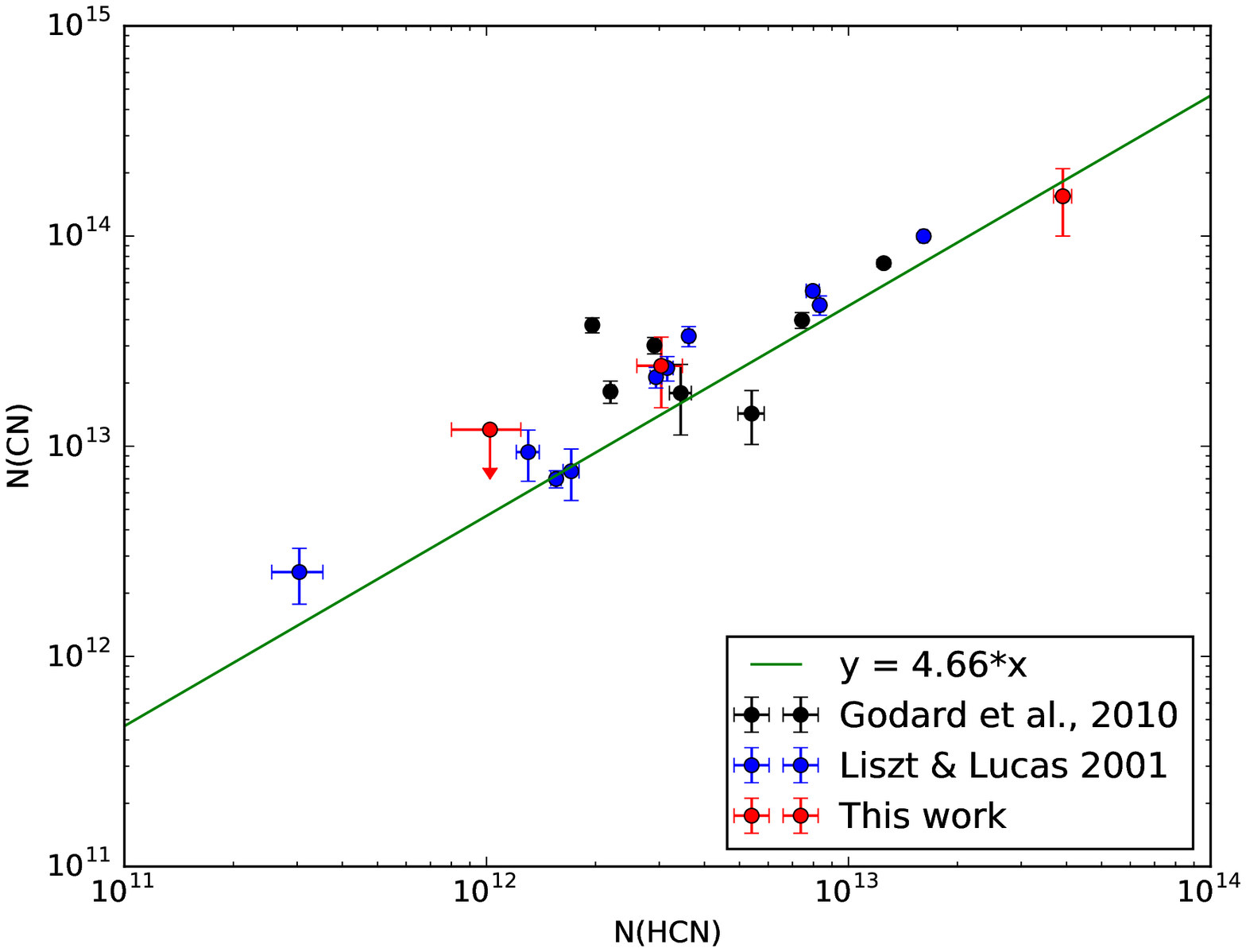}
\includegraphics[width=0.99\hsize]{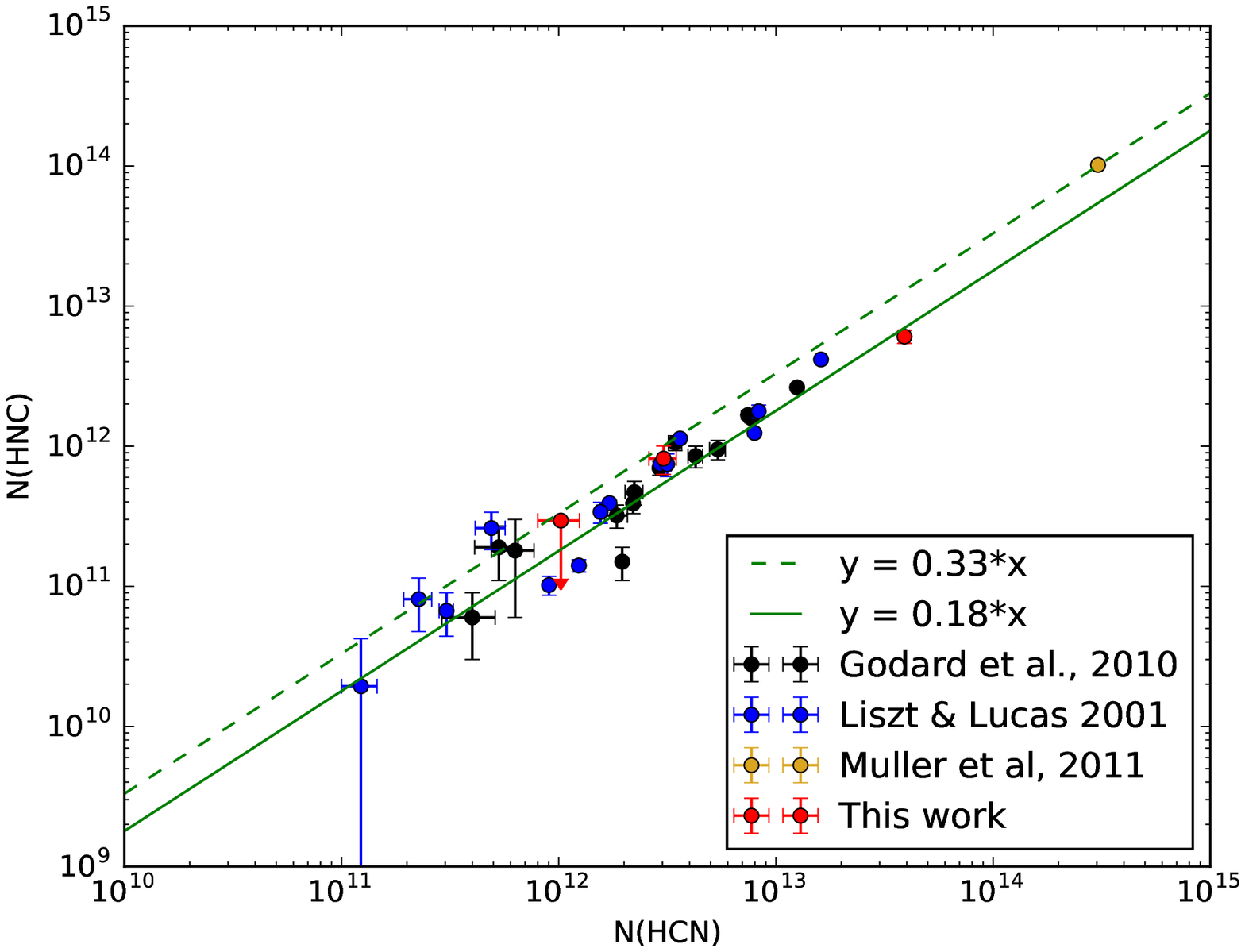}
}
\caption{Comparison of the column densities of CN, HCN, and HNC (the cyanogen- or CN-bearing molecular family). The black symbols correspond to the absorption study against Galactic star forming regions by \citet{Godard_et_al_2010}, the blue symbols correspond to the absorption study against quasars by \citet{Lucas_Liszt_2000}, yellow points refer to the absorption line study of a z=0.89 galaxy by \citet{Muller_et_al_2011}, and the red symbols correspond to this work. The red points with highest column density correspond to the gas in the Galactic disk ($v\sim 0 $ km s$^{-1}$), and the red-points in the middle correspond to the velocity component of $v\sim -153$ km s$^{-1}$. The lower red-points refer to the velocity component of $v\sim -192$ km s$^{-1}$}
\label{CN_HCN-HNCfamily}
\end{figure}

The comparison between the different families shows a high degree of correlation but a non-linear behavior.
Similar results were obtained by \citet{Godard_et_al_2010} in the absorption study toward star-forming regions in a large range of Galactic longitude along the Galactic plane.  Figure \ref{CN_HCN-HNCfamily} shows the column density comparison for the family of the cyanogen molecules. We include the data from  \citet{Liszt_Lucas_2001}, \citet{Godard_et_al_2010}, \citet{Muller_et_al_2011}, and the present work. We can see that our work  closely follows the relationship previously established by observations of absorption against quasars at high Galactic latitude and also by observations close to the Galactic plane in star formation regions. This global trend may indicate that the chemistry that drives the cyanogen-molecule family is similar in the three different environments. Note that the absorption lines against the quasar PKS 1830-211 from the z = 0.89 foreground galaxy \citet[][see the lower panel of Fig. 4]{Muller_et_al_2011} show higher column densities in all molecules by more than one order of magnitude than the other works. The green dashed line shows the fit including all data, and the continuous line shows the fit including only Galactic data (excluding the z = 0.89 absorption against the quasar PKS 1830-211). 

\begin{figure}
\centering
\vbox{
\includegraphics[width=0.99\hsize]{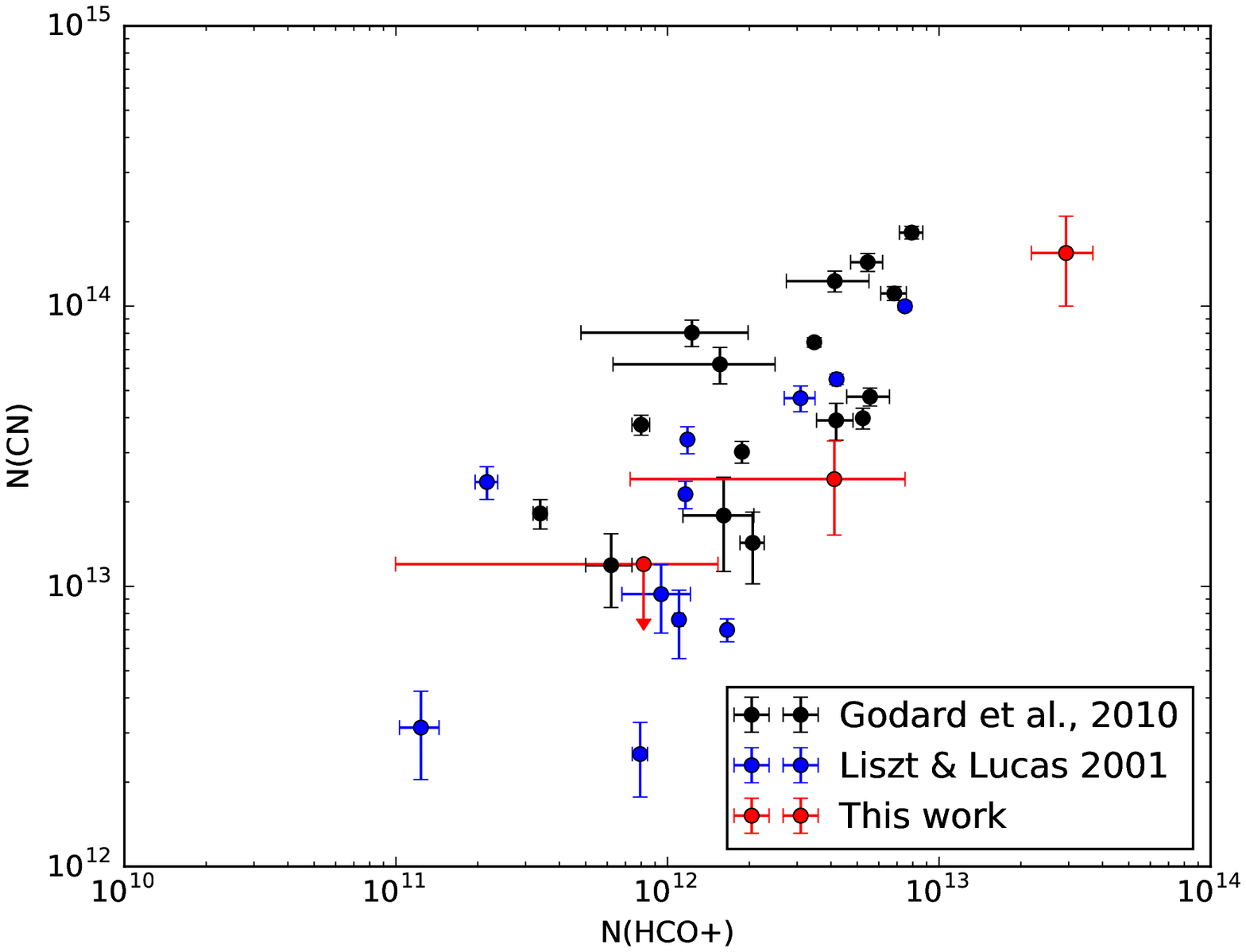}
\includegraphics[width=0.99\hsize]{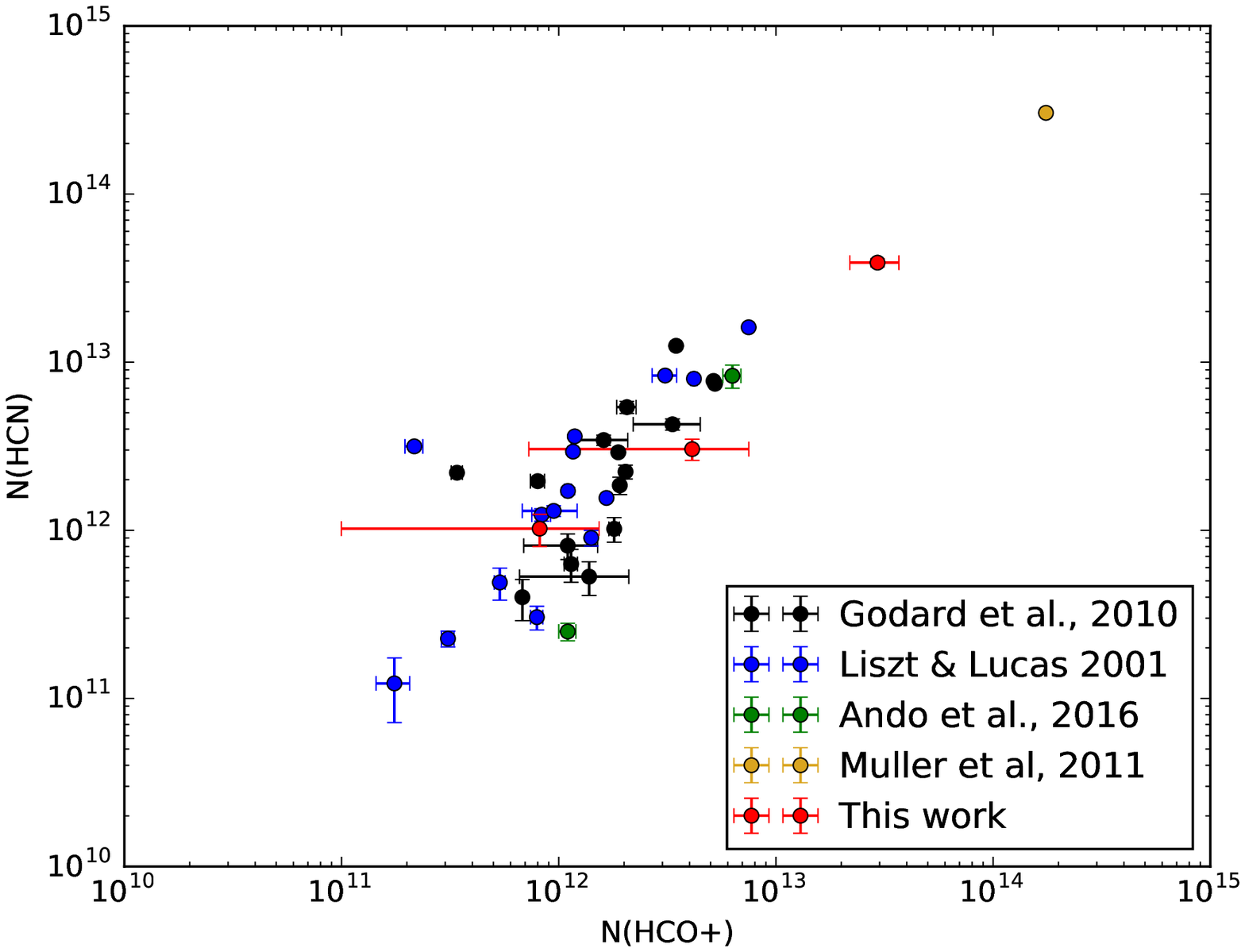}
\includegraphics[width=0.99\hsize]{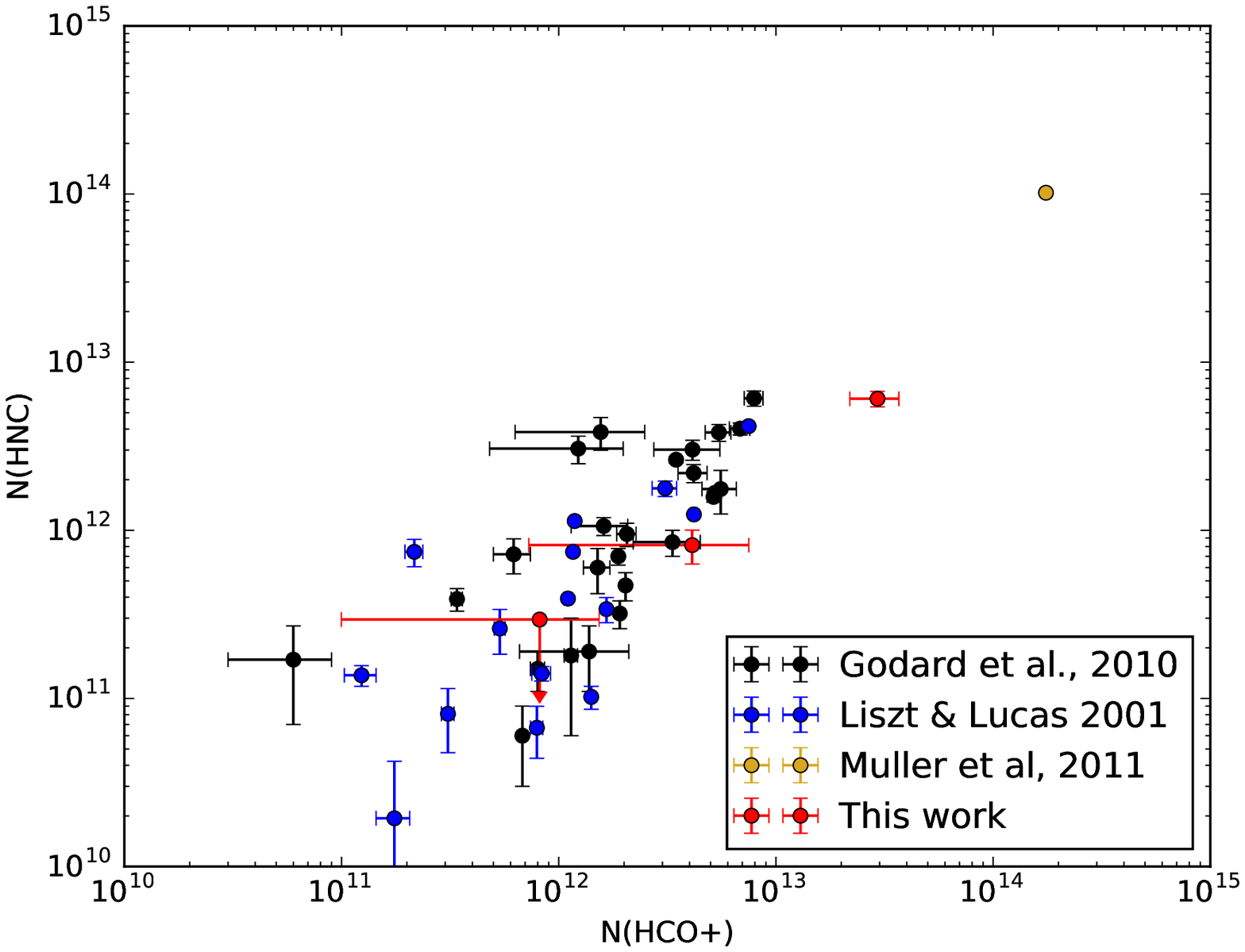}
}
\caption{Comparison of the column densities of the CN-bearing molecular family with HCO$^+$. The black symbols correspond to the absorption study against Galactic star forming regions by \citet{Godard_et_al_2010}, the blue symbols correspond to the absorption study against quasars by \citet{Lucas_Liszt_2000}, and the red symbols refer to this work. We also include in green symbols the recent work by \citet{Ando_et_al_2016} using Galactic ALMA calibrators and in yellow the absorption line study of a z=0.89 galaxy by \citet{Muller_et_al_2011}.}
\label{CN-HCO+family}
\end{figure}

\begin{figure}
\centering
\vbox{
\includegraphics[width=0.99\hsize]{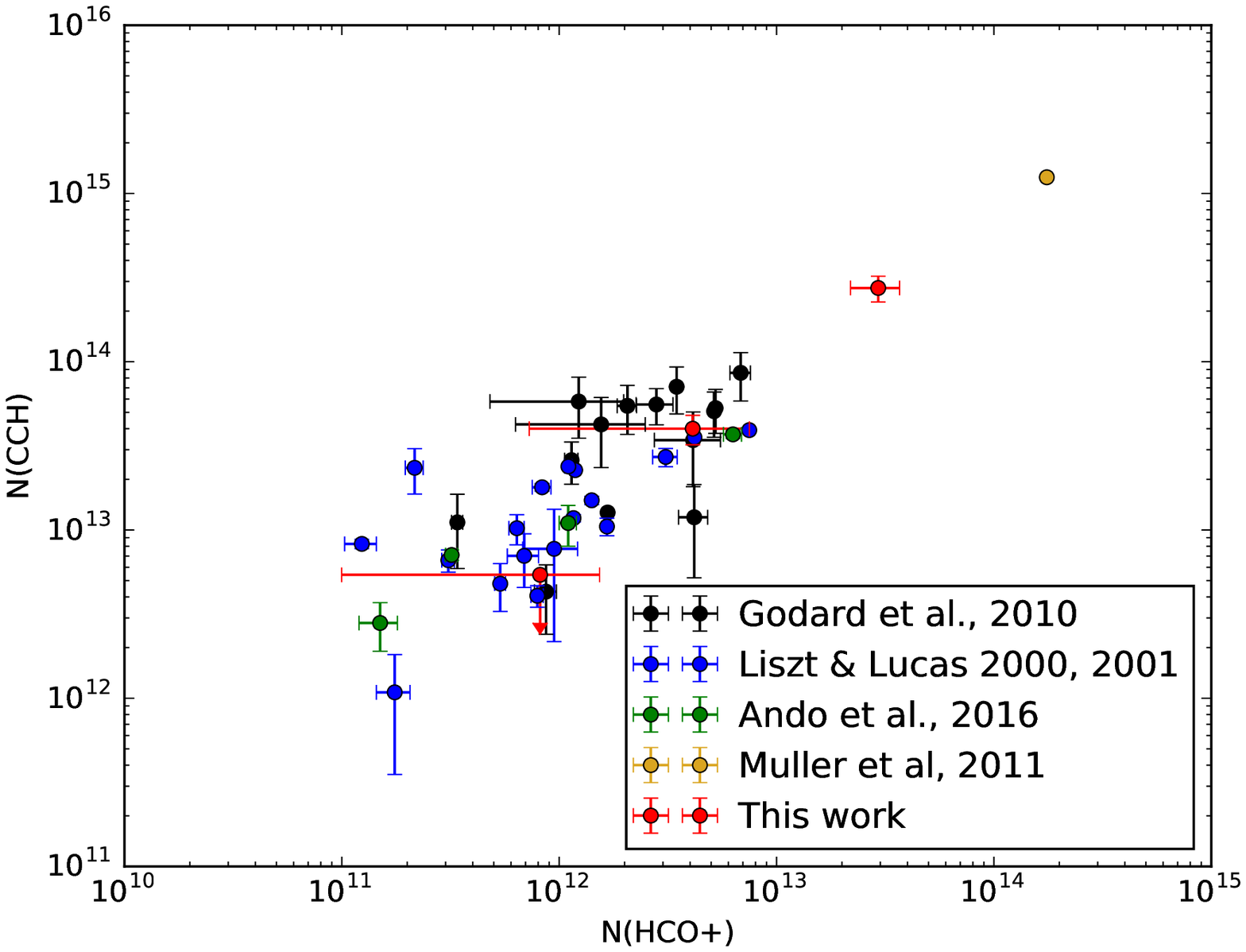}
\includegraphics[width=0.99\hsize]{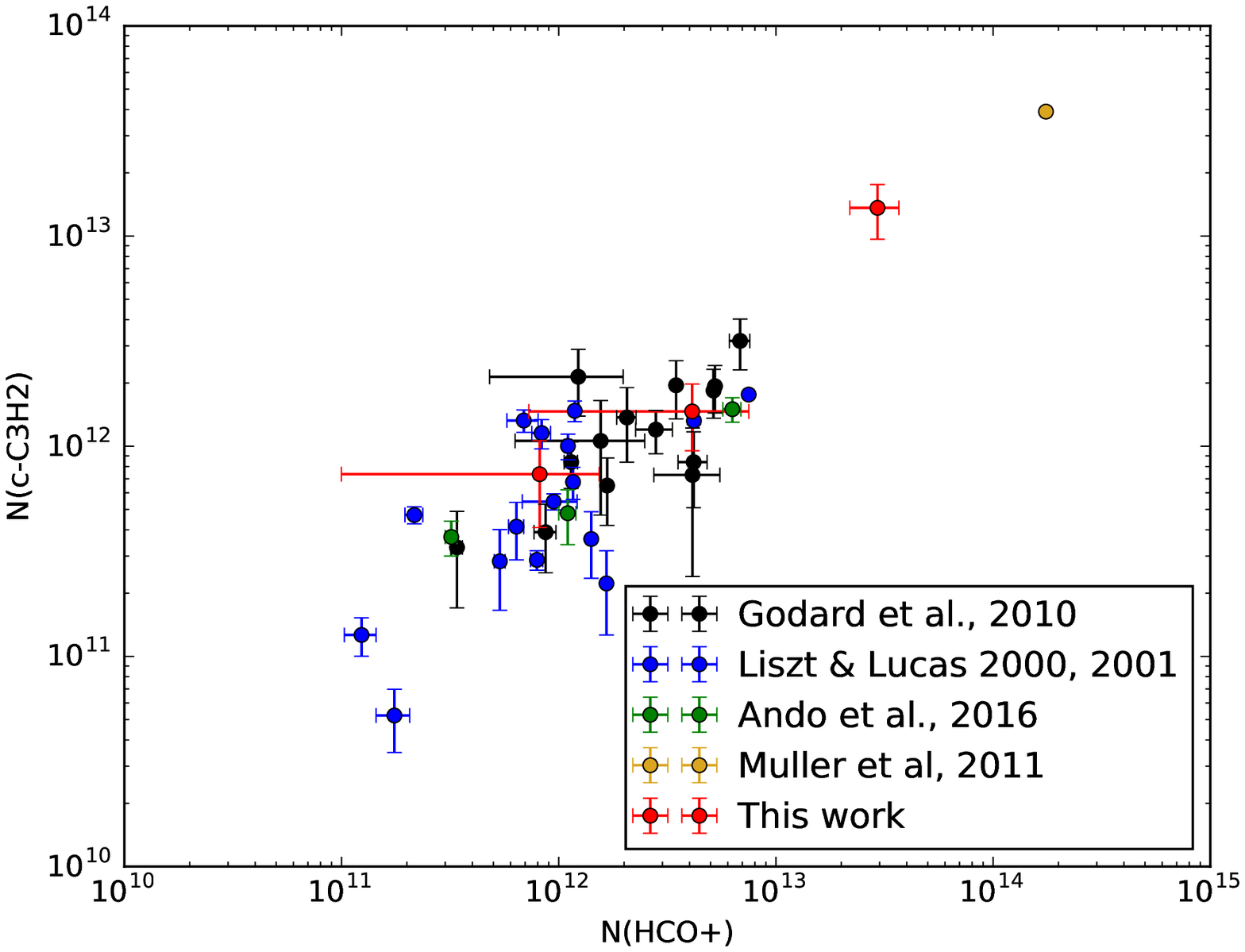}
}
\caption{Comparison of the column densities of the molecular C$_n$H$_m$-family with HCO$^+$. The black symbols correspond to the absorption study against Galactic star forming regions by \citet{Godard_et_al_2010}, the blue symbols correspond to the absorption study against quasars by \citet{Lucas_Liszt_2000}, and the red symbols refer to this work. We also include in green symbols the work by \citet{Ando_et_al_2016} and in yellow the line absorption study of the z=0.89 galaxy by \citet{Muller_et_al_2011}.}
\label{CnHm-family}
\end{figure}

In Fig. \ref{CN-HCO+family} we plot the cyanogen-molecule family against HCO$^+$ and we find that the HCO$^+$ column density is slightly larger than in previous works when compared with CN. We also include the recent work by \citet{Ando_et_al_2016} who detected molecular absorption lines of Galactic origin toward four radio-loud quasars which were observed as bandpass and complex gain calibrators in ALMA. Fig. \ref{CnHm-family} shows the column density of the C$_n$H$_m$-family against HCO$^+$. The data set of \citet{Godard_et_al_2010} was complemented by the CCH and c-C$_3$H$_2$ data from \citet{Gerin_et_al_2011}. The previous works also compared the linewidth of the different molecules, but this analysis is not possible with our data because of the lower spectral resolution of our survey ($\sim 10$ km s$^{-1}$). From Figs. \ref{CN_HCN-HNCfamily}, \ref{CN-HCO+family}, and \ref{CnHm-family} we can see that all data are well correlated which is an indication that the chemistries of the diffuse gas in the local ISM traced by the sample of \citet{Liszt_Lucas_2001}, the diffuse gas in the Galactic plane traced by \citet{Godard_et_al_2010}, and the diffuse gas in the Galactic bulge of this work are similar.  \citet{Godard_et_al_2010} showed that the column densities of the different molecules cannot be reproduced by UV-dominated chemical models (PDR models) but by dissipation of turbulent energy models. Further observations of higher rotational transitions with higher spectral resolution are needed to check our assumption of $T_{ex}=T_{bg}$ and to derive physical properties like kinetic temperature and volume density of this diffuse gas and to compare this with chemical models.

\subsection{Mass of the low density gas and comparison with gas in emission in the CMZ \label{amountgas}}
Table \ref{abundanceCMZ} shows typical column densities along several lines-of-sight towards the CMZ for the detected molecules in this work. The column density of the gas detected in absorption in the bulge of the Galaxy from this work is 1-3 orders of magnitude lower than that of gas in emission in the CMZ.
To derive reliable fractional abundances for the detected species in this survey, we need to compute molecular column densities N(H$_2$). From the available data of this work it is not possible to derive N(H$_2$). However, we can estimate  N(H$_2$) using the relation shown in Fig. 3 of \citet{Liszt_Lucas_2001} between CN and H$_2$ derived by optical observations of \citet{Federman_et_al_1994}. The relation between N(CN) and N(H$_2$) is non-linear with a power-law slope of 1.31.  Using this relation with our data, we can infer a molecular hydrogen column of $1.12\times 10^{22}$ cm$^{-2}$ for the potentially local gas (0 km s$^{-1}$), $2.71\times 10^{21}$ cm$^{-2}$  for the velocity component at $-153$ km s$^{-1}$, and an upper limit ($3\sigma$) of $1.60\times 10^{21}$ cm$^{-2}$ for the velocity component at $-192$ km s$^{-1}$.
These results are in good agreement with the recent work by \citet{Gerin_Liszt_2017}, who derived N(H$_2)=1.21\times 10^{21}$ cm$^{-2}$ and $2.86\times 10^{20}$ cm$^{-2}$ for the velocity components at $\sim -150$ km s$^{-1}$ and $\sim -190$ km s$^{-1}$ respectively, using N(HCO$^+$)/N(H$_2$)$=3\times 10^{-9}$ (the local value for solar metallicity). If we use this HCO$^+$ abundance value, we derive a N(H$_2)=1.37\times 10^{21}$ cm$^{-2}$ and $2.72\times 10^{20}$ cm$^{-2}$ for the velocity components at $\sim -150$ km s$^{-1}$ and $\sim -190$ km s$^{-1}$ respectively. Riquelme et al., 2017 (submitted) estimated HCO$^+$ abundances of $>3.8\times 10^{-9}$ for several clouds in the CMZ, which yields similar N(H$_2$) as the one derived using the non-linear relation between CN and H$_2$ and the HCO$^+$ abundance value used by \citet{Gerin_Liszt_2017}.

We can obtain a very rough estimate of the total mass of the low density gas, using the column density of the velocity component at $-153$ km s$^{-1}$. Since the quasar is located at  $(l,b)=(-2{^\circ}.13,-1{^\circ}$), we assume that the low density gas has, at least, an extension of 4${^\circ}$.3 in Galactic longitude if this component would be symmetrically distributed \footnote{This is not actually true for the molecular gas seen in emission in the CMZ. The CMZ shows an asymmetry in the distribution of the molecular gas with 2/3 of the dense gas being placed at positive longitudes \citep{Morris_Serabyn_1996}, while our source is located at negative longitude} and would show a spherical geometry, with  a radius equal to the projected distance between our source and the GC. Considering a distance of 8.5 kpc for the GC, the total mass of the diffuse component could be as high as $1.9\times 10^{7}$ M$_{\odot}$.
In the literature there are many independent estimations of the total gas mass for the GC \citep[see, ][for a review]{Ferriere_et_al_2007}. For example, \citet{Launhardt_et_al_2002} derive a total mass of the interstellar hydrogen for the entire CMZ of $\sim 6\times 10^7$ M$_{\odot}$. 
The molecular gas observed in emission in the CMZ shows different components, one dense and cool (T$_{\rm kin}<50$ K, n$\sim 10^{3.5}$ cm$^{-3}$), and the n$\sim 10^{4}$ cm$^{-3}$, T$_{\rm kin}>50$ K component of \citet{Ao_et_al_2013,Ginsburg_et_al_2016} and also a warmer and thinner component with kinetic temperature $\sim 150$ K and n$\sim 10^{2.5}$ cm$^{-3}$. \citet{Dahmen_et_al_1998} estimate $1.2-6.4\times 10^7$ M$_{\odot}$ for the dense gas  and $0.7-1.4 \times 10^7$ M$_{\odot}$ for the thin gas for the central $\sim 600$ pc of the GC.
Although we only have one line-of-sight, this result may indicate that the low density gas found in this work represents a considerable fraction of about 1/3 of the total gas mass in the GC.

\begin{table*}
\caption{Column densities for the detected species (in emission) in the GC along different lines-of-sight towards the CMZ from recent previous works.}             
\label{abundanceCMZ}      
\centering                          
\begin{tabular}{c c c c c c c}        
\hline\hline                 
Molecule      & SgrB2N$^a$ & SgrB2M$^a$ & CND$^b$    & LOS+0.693$^c$ & LOS-0.11$^c$  & CMZ$^{d, e}$ \\    
	      & [cm$^{-2}$]& [cm$^{-2}$]& [cm$^{-2}$]&[cm$^{-2}$]& [cm$^{-2}$] & [cm$^{-2}$]\\
\hline  
c-C$_3$H$_2$  & $2.50\times 10^{14}$           & $2.60\times 10^{14}$          &            & $9.4 \pm 2.5 \times 10^{13}$          &  $6.0 \pm 0.7 \times 10^{13}$      &     \\
CCH           & $2.00\times 10^{15}$           & $5.00\times 10^{15}$          &            & $3.162 \pm 0.186 \times 10^{15}$          & $2.512 \pm 0.24 \times 10^{15}$      &      \\
HCN	      & $1.07\times 10^{17}$           & $1.79\times 10^{16}$         & $3.38 \pm 0.40 \times 10^{15}$    & $>2.40 \times 10^{15}$         &   $>4.80 \times 10^{15} $   &  $1.7 \times 10^{15} $    \\
HCO$^+$       & $5.07\times 10^{15}$           & $7.45\times 10^{15}$          & $5.26 \pm 1.28 \times 10^{14}$    & $5.70 \pm 0.13 \times 10^{14}$         & $3.6 \pm 0.08 \times 10^{14}$      &  $5.0 \times 10^{14}$     \\
HNC 	      & $4.79\times 10^{15}$           & $5.98\times 10^{15}$          & $1.90 \pm 0.29 \times 10^{14}$          & $>2.4\times 10^{15}$          &  $>1.8\times 10^{15}$      &   $7.0 \times 10^{14}$   \\
CS            & $2.59\times 10^{17}$           & $1.49\times 10^{16}$          & $1.07 \pm 0.13 \times 10^{15}$    &           &              \\
$^{13}$CO     & $3.08\times 10^{18}$           & $3.45\times 10^{18}$          &            & $2.81 \pm 0.0644 \times 10^{17}$   & $1.15 \pm 0.095 \times 10^{17}$    &  \\
CN            & $6.93\times 10^{14}$           & $9.90\times 10^{15}$          & $5.48 \pm 3.75 \times 10^{15}$             & $1.698 \pm 0.174 \times 10^{15}$ & $1.738 \pm 0.195 \times 10^{15}$     &       \\
\hline                                     
\end{tabular}
\tablefoot{ $^a$ \citet{Belloche_et_al_2013} of the velocity component closest to the systemic velocity of Sgr B2N (64 km s$^{-1}$) and Sgr B2M (62 km s$^{-1}$), $^b$ \citet{Harada_et_al_2015}, $^c$ \citet{Armijos-Abedano_et_al_2015}, $^d$ \citet{Jones_et_al_2012}, $^e$ average value over the CMZ corrected for optical depth effects. }
\end{table*}

\subsection{Origin of the excitation of the diffuse gas in the bulge and its relationship with the CMZ}

There are still many unknown parameters in the low excitation absorption gas found in this work, referring to its relationship with the CMZ and the association with the warm low density component found by infrared absorption spectroscopy of H$_3^+$ and CO by \citet{Oka_et_al_2005}, \citet{Goto_et_al_2008} and \citet{Geballe_Oka_2010}. The increased amount of sight-lines available for absorption studies in H$_3^+$ and CO \citep{Geballe_et_al_2015} show that the warm, low density  gas fills a large fraction of the volume of the CMZ, but they do not cover regions outside the CMZ associated with the bulge of the Galaxy.
\citet{Goto_et_al_2014} observed the H$_3^+$ and ro-vibrational transitions of CO on sightlines towards two luminous infrared sources located in the Central Stellar Cluster associated with the GC and found that the absorption occurs in three kinds of gaseous environments: 1) cold dense and diffuse gas associated with foreground spiral arms containing both species (H$_3^+$ and CO); 2) warm and diffuse gas absorbing over a wide and mostly negative velocity range ($-180$ to $+20$ km\,s$^{-1}$), which may be filling a large fraction of the CMZ, containing H$_3^+$ and little CO; and 3) warm, dense and compact clouds which contain both species. 
With respect to the second kind of gas, \citet{Goto_et_al_2014} and \citet{Goto_et_al_2015} relate the large negative velocities ($-160$ to $-110$ km\,s$^{-1}$) to the Expanding Molecular Ring \citep[EMR, ][]{Scoville_et_al_1972}, which may be expanding from the GC with a radial velocity of $\sim 200$ km\,s$^{-1}$. As examples of the third kind, \citet{Goto_et_al_2014} find absorption along lines-of-sight towards the Central Cluster at velocities $\sim $50 km\,s$^{-1}$, which may be associated with an inward extension of the circunnuclear disk.
Referring to the clouds detected in this work, their relationship with the diffuse and warm gas in the CMZ is not clear. Also not settled is the question whether they are associated with features seen in emission in the GC region, such as EMR, or if they are warm, dense and compact clouds, similar to the third kind of environment discussed above which is unlikely since they could be also detected in emission. Higher rotational transitions of the detected molecules are needed to derive the physical properties of the detected clouds, as well as infrared absorption studies towards larger regions of the GC to unveil the extension of the warm and diffuse gas found by \citet{Oka_et_al_2005}.

\subsection{A new line of sight for Galactic diffuse ISM studies}
The velocity component at v$\sim 0$ km s$^{-1}$ adds a new line-of-sight for Galactic diffuse ISM studies. At the moment, there is only a limited number of Galactic absorption systems known in the literature ($\sim$ 30) and only few of them have been extensively studied \citep{Lucas_Liszt_1996, Lucas_Liszt_2000, Liszt_Lucas_2001, Liszt_et_al_2014}. Recently, \citet{Ando_et_al_2016} using calibrators in the ALMA Archive found 4 absorption systems of Galactic origin (three of them are new detections). 
In this work, the detected absorption against the quasar J1744-312 at the velocity of v$\sim 0$ km s$^{-1}$ is of Galactic origin, likely coming from local gas or gas belonging to the spiral arms. Radial components of rotation velocities collapse to zero in the direction of the Galactic center. This velocity component shows a rich chemistry with the detection of the 12 species listed in Table \ref{table:2}, and in general, its column density is about one order of magnitude higher than the velocity components corresponding to the bulge components with highly negative velocities. From these species, the detection of HCO at the velocity component $\sim 0$ km\,s$^{-1}$ is particularly remarkable, which is the sixth detection in absorption found up to date in the diffuse medium \citep{Liszt_et_al_2014, Ando_et_al_2016}. HCO is a PDR tracer even in the diffuse medium ($n_{H_2}\sim10^1-10^2$ cm$^{-3}$), and its detection is the result of the ionization of carbon (C$^+$) by distant OB stars, which together with H$_2$ lead to the formation of CH$_2$, and then HCO, probably in the gas phase through the reaction ${\rm O+CH_2 \rightarrow HCO + H}$ \citep{Gerin_et_al_2009}. In this work, the column density ratio $N({\rm HCO})/N({\rm H^{13}CO^+})$ commonly used as diagnostic of the presence of far-UV radiation field has a value of 9, which is higher than 1  \citep[the threshold for a PDR-like environment,][]{Gerin_et_al_2009}, but lower than the values derived from previous absorption studies (19.3) and towards well-known PDRs.

We also detected SiO at the velocity component $\sim 0$ km\,s$^{-1}$ with a low fractional abundance of N(SiO)/N(H$_2$) of $1\times 10^{-10}$ consistent with the diffuse  and translucent clouds studied using extragalactic continuum sources by \citet{Lucas_Liszt_2000b}. SiO has been extensively studied in the molecular clouds of the GC region finding high abundances associated with large scale shocks \citep[see e.g., ][among others]{Martin-Pintado_et_al_1997, Huettemeister_et_al_1998,  Riquelme_et_al_2010b, Mihn_et_al_2015,Tsuboi_et_al_2015}, and its enhancement in the disk has been associated to outflows. SiO can also be formed without a shock at a low abundance level as shown by previous detections in photon dominated regions \citep{Schilke_et_al_2001} and in diffuse gas \citep{Lucas_Liszt_2000b}, and its underabundance in our case could be due to         a high degree of depletion of Si from the gas phase.


\section{Conclusions}

   \begin{enumerate}
   \item Using ALMA in the 3 mm wavelength range, we have detected a low excitation gas component related to the bulge of the Galaxy, seen in absorption against the background quasar J1744-312. In addition, we also find near zero velocity absorption lines that may originate from the Galactic disk.
   \item If we assume that the derived column density from the bulge is uniform over the volume subtended by the source, we tentatively estimate that the amount of this low excitation gas in the bulge may be as high as $1.9\times 10^7$ M$_{\odot}$, which in such scenario would corresponds to a fraction of 1/3 of the total gas of the CMZ.
   \item The comparison of the column densities derived in this work with previous studies indicate that the chemistry of the diffuse gas in the local ISM, Galactic plane, and bulge of the Galaxy may be similar. This result needs to be confirmed by detailed observations of higher rotational transitions with higher spectral resolution to derive reliable column densities and physical properties of this gas.
   \end{enumerate}

\begin{acknowledgements}
This work was carried out within the Collaborative Research Council 956, subproject A5, funded by the Deutsche Forschungsgemeinschaft (DFG). L.B. and R.F. acknowledge support by CONICYT CATA-Basal project PFB-06. We thank to Arnaud Belloche for useful discussions. We thank to Holger S. P. M\"uller for providing the HCO partition function value for 2.725 K. We thanks to the anonymous referee for his/her critical reading and constructive comments that helped to improve this manuscript.
\end{acknowledgements}

\bibliographystyle{aa} 
\bibliography{references_CMZ} 
\clearpage

\begin{appendix}
\section{Complementary tables and figures}
\begin{table*}
\caption{Molecular data for detected transitions}             
\label{table:1}      
\centering                          
\begin{tabular}{c c c c c c c c c}        
\hline\hline                 
Molecule & Transitions & Frequency & $\mu$  & Q$_{2.73}$ & A$_{ul}$  & E$_l$ & Reference\\    
	 &	       & [GHz]		& [Debye]&        & [s$^{-1}$]   & [cm$^{-1}$]         &     \\
\hline                        
c-C$_3$H$_2$& 2$_{1,2}-1_{0,1}$& 85.338894  &  3.27 & 8.8732$^b$& 2.3221$\times 10^{-5}$& 1.6332 &CDMS \\
H$^{13}$CN & 1-0, F=1-1,2-1,0-1& 86.339921  & 2.9852&  5.1261 & 2.2255$\times 10^{-5}$& 0.0 & CDMS \\		
HCO& $1_{0,1}-0_{0,0}$, J=3/2-1/2, F=2-1&86.67076& $\mu_a$=1.3626,$\mu_b$=0.700 & 6.7785$^c$& 4.6889$\times 10^{-6}$&  0.0129 & JPL\\
HCO& $1_{0,1}-0_{0,0}$, J=3/2-1/2, F=1-0&86.70836& $\mu_a$=1.3626,$\mu_b$=0.700 & 6.7785$^c$& 4.5951$\times 10^{-6}$&  0.0000 &  JPL\\
H$^{13}$CO$^+$&	1-0	      &86.7542884&3.90 & 1.7028 & 3.8535$\times 10^{-5}$& 0.0  & CDMS \\
HCO& $1_{0,1}-0_{0,0}$, J=1/2-1/2, F=1-1&86.77746& $\mu_a$=1.3626,$\mu_b$=0.700 & 6.7785$^c$& 4.6065$\times 10^{-6}$&0.0129&JPL\\
HCO& $1_{0,1}-0_{0,0}$, J=1/2-1/2, F=0-1&86.80578& $\mu_a$=1.3626,$\mu_b$=0.700 & 6.7785$^c$& 4.7124$\times 10^{-6}$&0.0129&JPL\\
SiO        &v=0, 2-1          & 86.846985   & 3.0882 & 2.9760  &  2.9274$\times 10^{-5}$&   1.4485 &CDMS\\
CCH & N=1-0, J=3/2-1/2, F=1-1 & 87.284105  & 0.770 & 6.7735 &2.5989$\times 10^{-7}$& 0.0015 &  CDMS \\
CCH & N=1-0, J=3/2-1/2, F=2-1 & 87.316898  & 0.770 & 6.7735 &1.5314$\times 10^{-6}$& 0.0015 &  CDMS \\
CCH & N=1-0, J=3/2-1/2, F=1-0 & 87.328585  & 0.770 & 6.7735 &1.2718$\times 10^{-6}$& 0.0      &  CDMS \\
CCH & N=1-0, J=1/2-1/2, F=1-1 & 87.401989  & 0.770 & 6.7735 &1.2749$\times 10^{-6}$& 0.0015 &  CDMS \\
CCH & N=1-0, J=1/2-1/2, F=0-1 & 87.407165  & 0.770 & 6.7735 &1.5361$\times 10^{-6}$& 0.0015 &  CDMS \\
CCH & N=1-0, J=1/2-1/2, F=1-0 & 87.446470  & 0.770 & 6.7735 &2.6133$\times 10^{-7}$& 0.0      &  CDMS \\
HCN	 & 1-0	       		   & 88.6316022   &  2.9852 &  5.03   & 2.4074$\times 10^{-5}$&  0.0 & CDMS \\
HCO$^+$  & 1-0	      		   & 89.1885247   &  3.90   &  1.6691 & 4.1869$\times 10^{-5}$&  0.0 & CDMS \\
HNC 	 & 1-0	       		   & 90.6635680   &  3.05   & 1.6497  & 2.6896$\times 10^{-5}$&  0.0 & CDMS \\
CS       & 2-1	       		   & 97.9809533   &  1.958  & 2.6827  & 1.6792$\times 10^{-5}$& 1.6342  & CDMS \\
$^{13}$CO& 1-0         		  & 110.2013543   & 0.11046 &  2.8912 & 6.3331$\times 10^{-8}$& 0.0  & CDMS \\
CN       & 1-0 1/2-1/2 F=1/2-3/2  & 113.1441573  &   1.45  & 8.5178  &  1.053$\times 10^{-5}$& 0.0    &    CDMS \\
         & 1-0 1/2-1/2 F=3/2-1/2  & 113.1704915  &         & 8.5178  & 5.1451$\times 10^{-6}$& 0.0007    &    CDMS \\
         & 1-0 1/2-1/2 F=3/2-3/2  & 113.1912787$^{a}$  &         & 8.5178  &  6.683$\times 10^{-6}$& 0.0   &    CDMS \\
         & 1-0 3/2-1/2 F=3/2-1/2  & 113.4881202  &         & 8.5178  & 6.7363$\times 10^{-6}$& 0.0007    &    CDMS \\
         & 1-0 3/2-1/2 F=5/2-3/2  & 113.4909702  &         & 8.5178  & 1.1924$\times 10^{-5}$& 0.0    &    CDMS \\
         & 1-0 3/2-1/2 F=1/2-1/2  & 113.4996443  &         & 8.5178  & 1.0629$\times 10^{-5}$& 0.0007    &    CDMS \\
         & 1-0 3/2-1/2 F=3/2-3/2  & 113.5089074  &         & 8.5178  & 5.1904$\times 10^{-6}$& 0.0    &    CDMS \\
\hline                                   
\end{tabular}
\tablefoot{$\mu$ corresponds to the dipole moment along the molecular axis.  Q$_{2.73}$ is the Partition Function at T$_{\rm ex}=2.73$ K. A$_{ul}$ is the Einstein coefficient. E$_l$ is the energy of the lower level. $^a$ Line used to estimate the optical depth (see Section \ref{fitting}). $^b$ Partition function for ortho c-C3H2. $^c$ Holger S. P. Muller, private communication.}
\end{table*}

\begin{figure*}
\centering
\vbox{
\hbox{
\includegraphics[width=0.5\hsize]{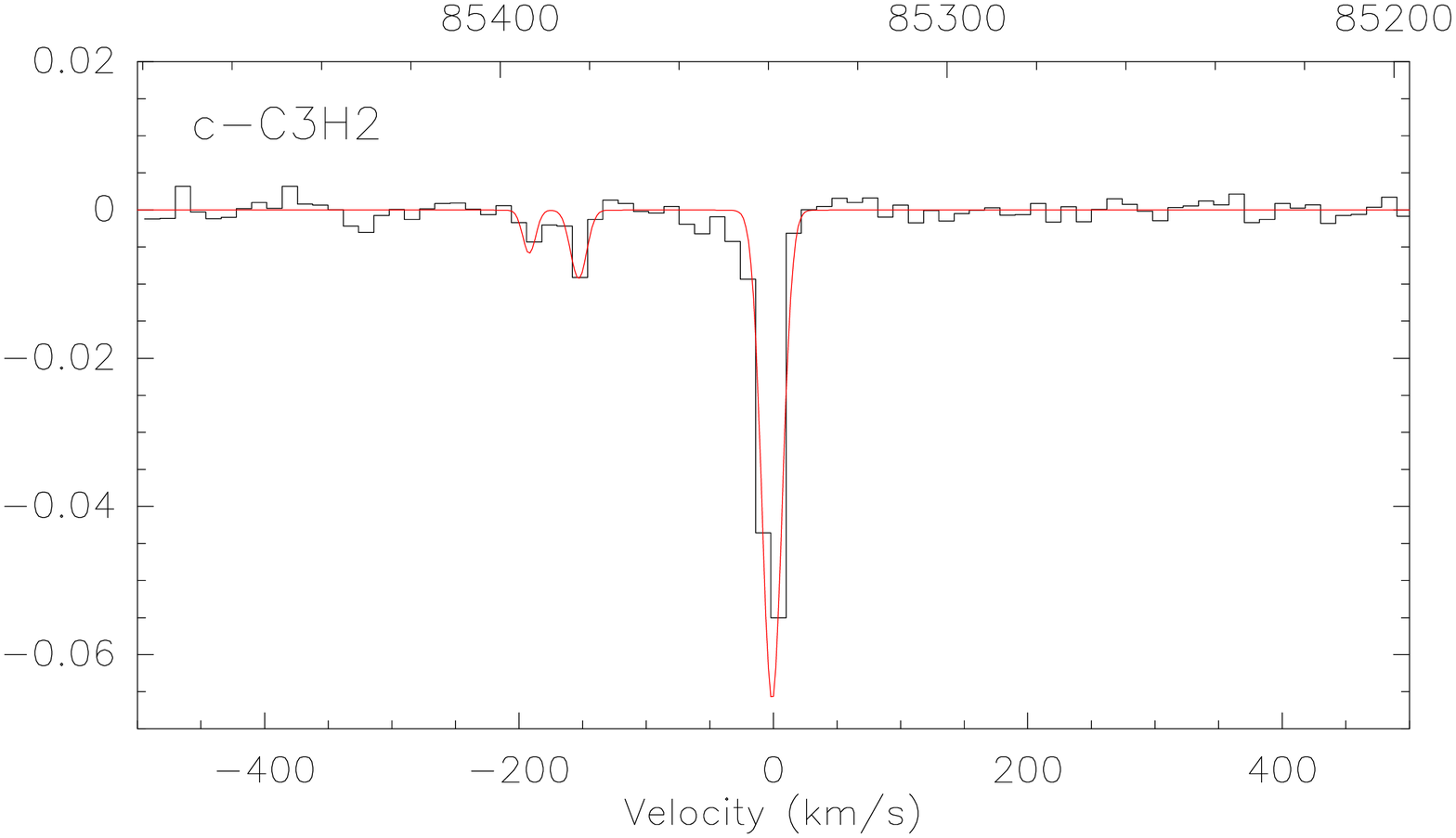}
\includegraphics[width=0.5\hsize]{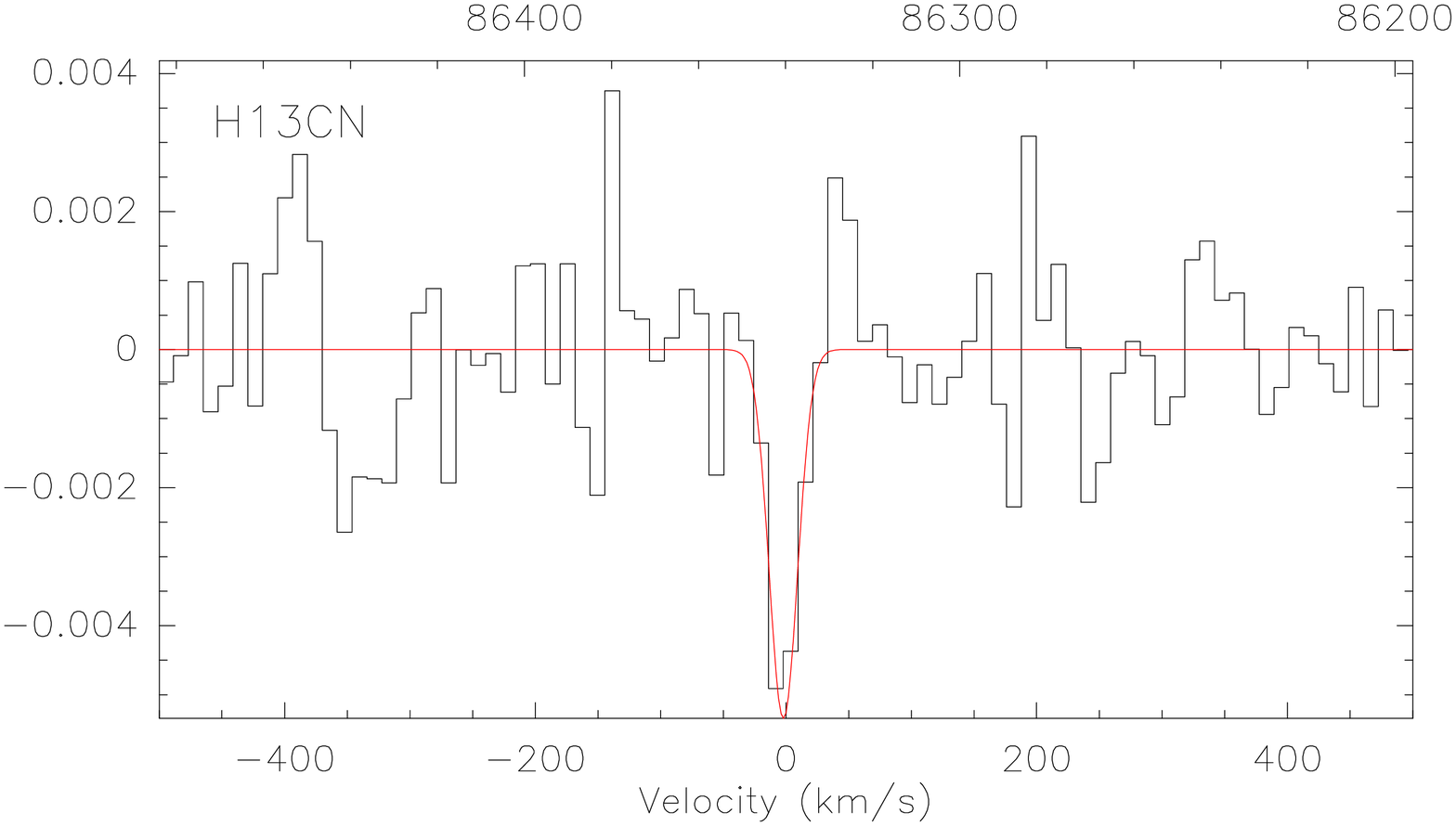}
}
\hbox{
\includegraphics[width=0.5\hsize]{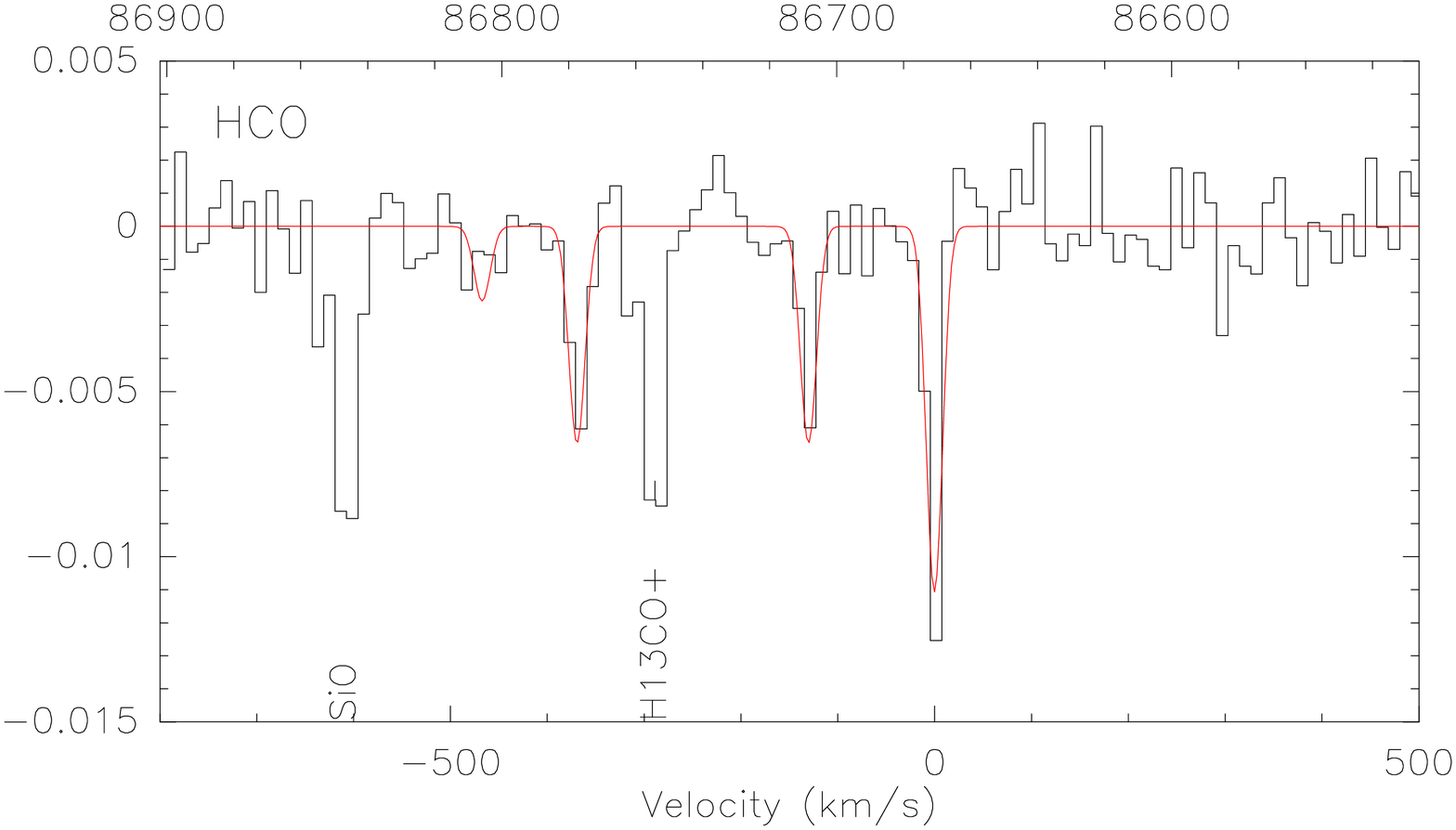}
\includegraphics[width=0.5\hsize]{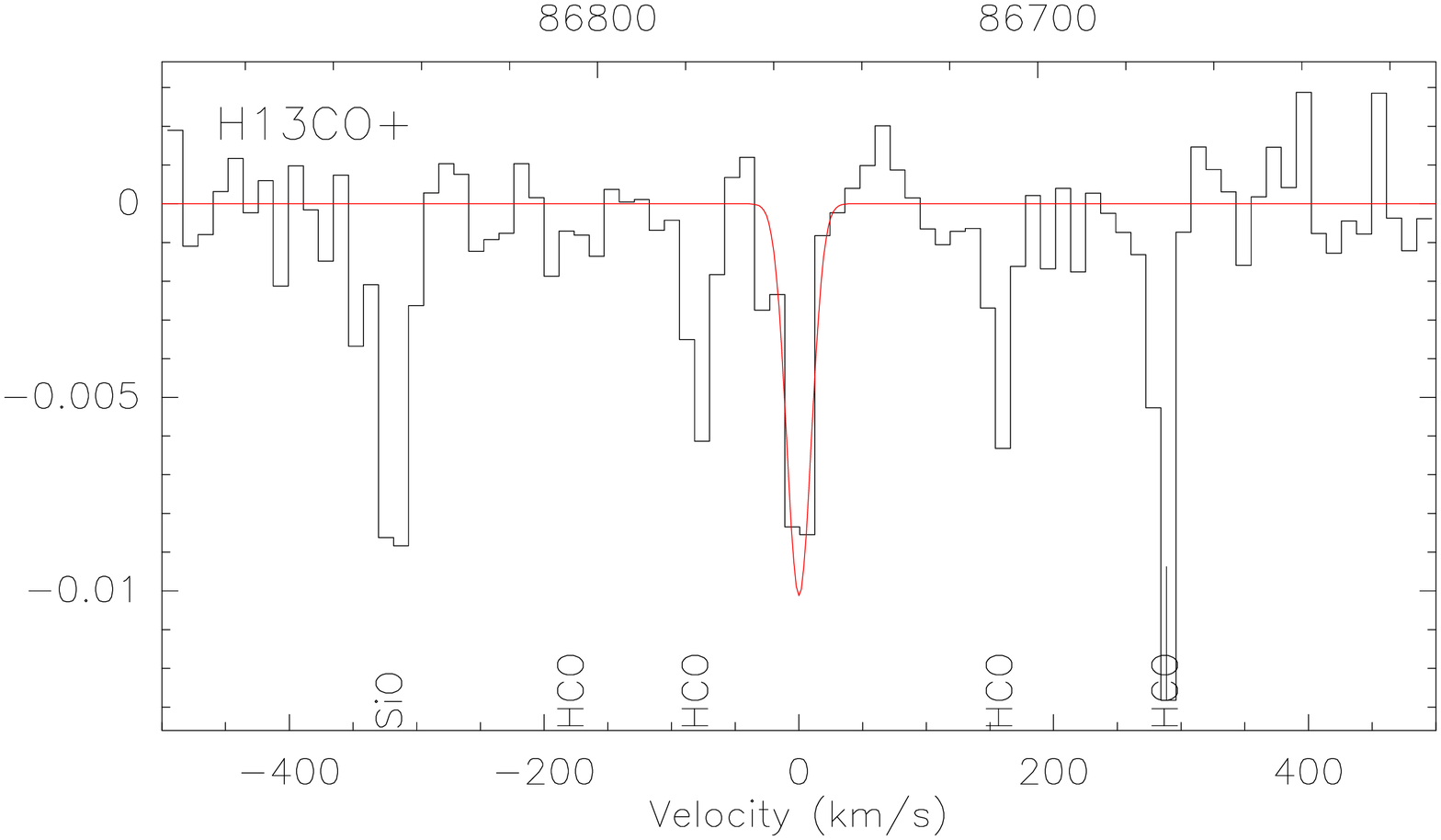}
}
\hbox{
\includegraphics[width=0.5\hsize]{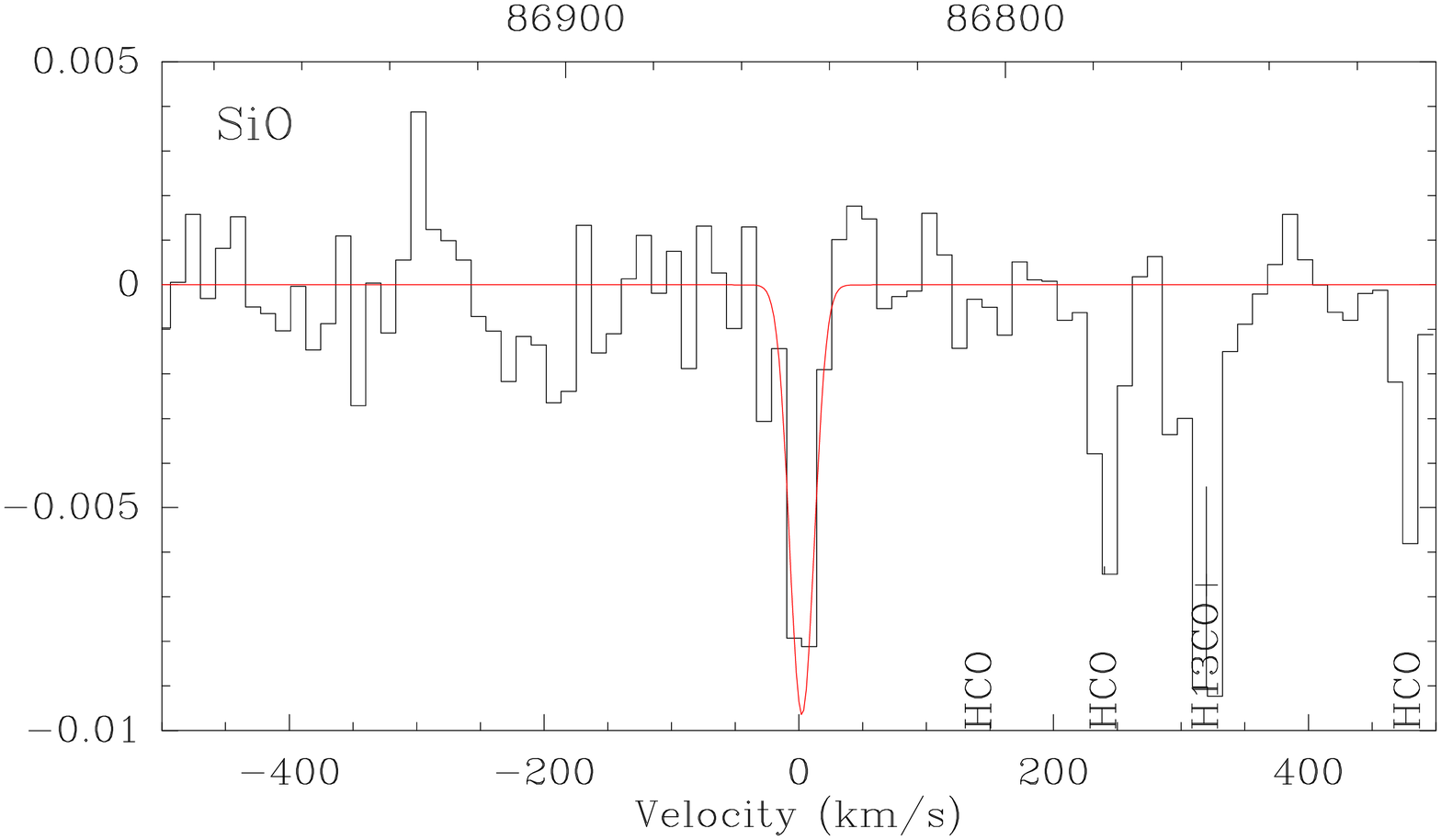}
\includegraphics[width=0.5\hsize]{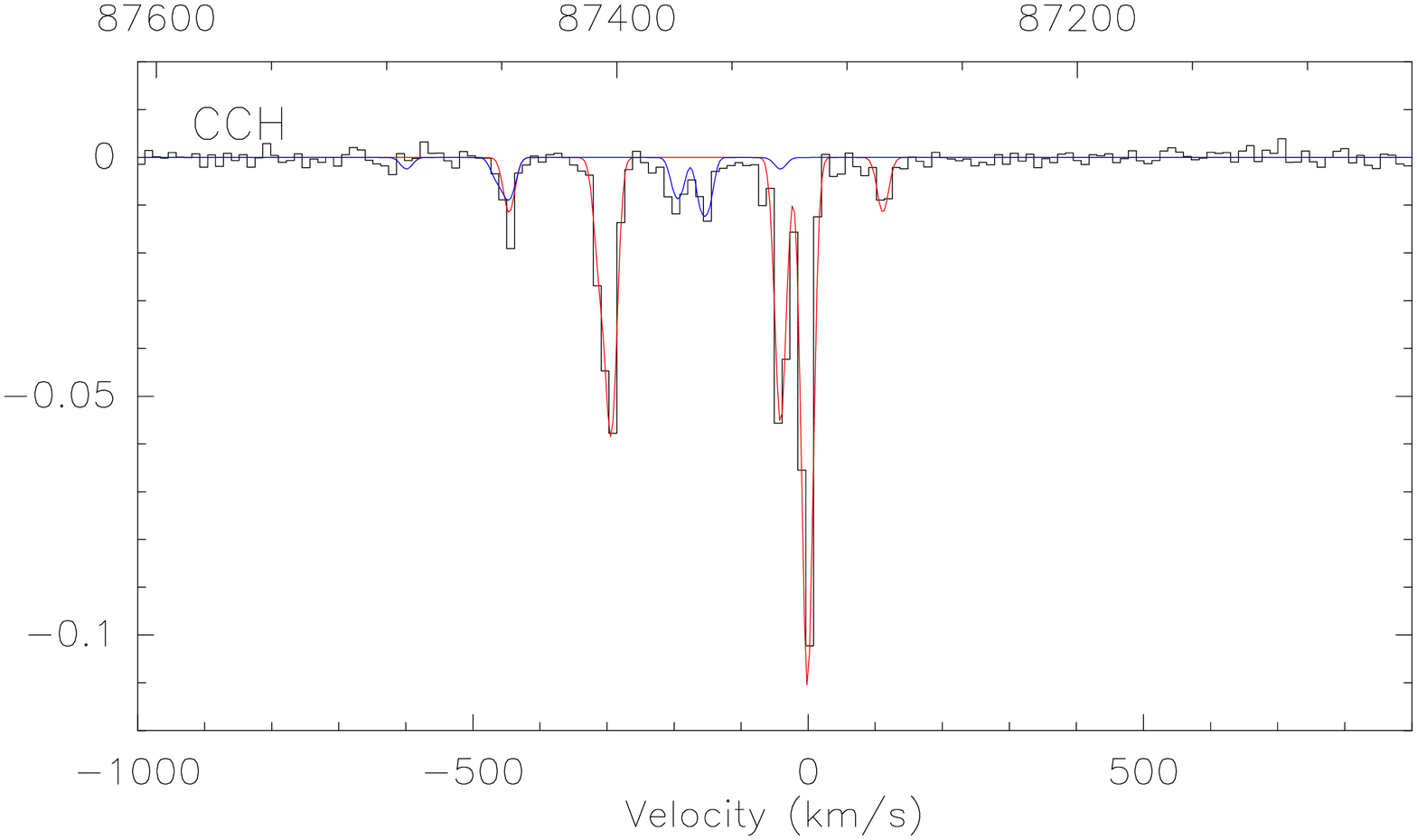}
}
\hbox{
\includegraphics[width=0.5\hsize]{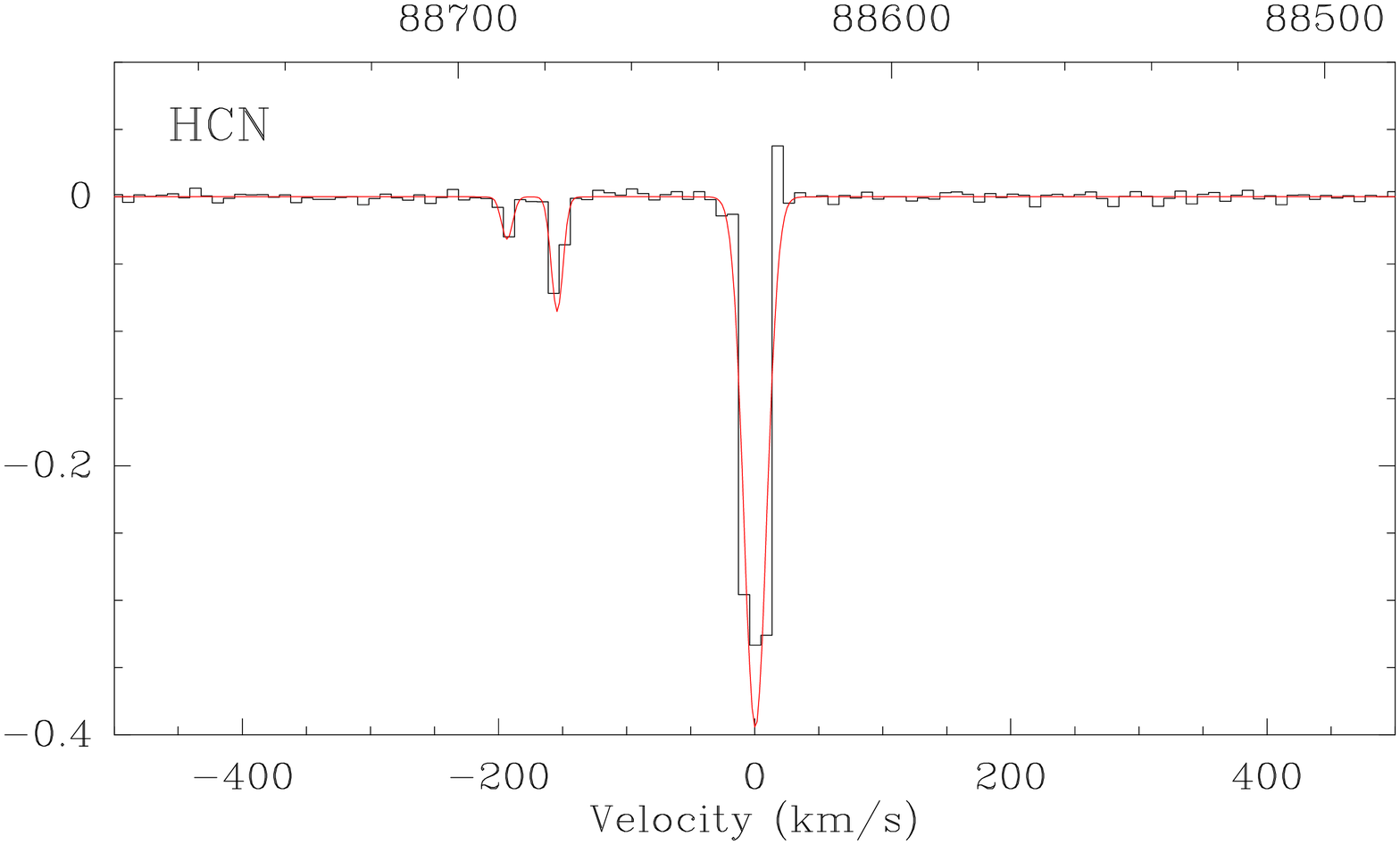}
\includegraphics[width=0.5\hsize]{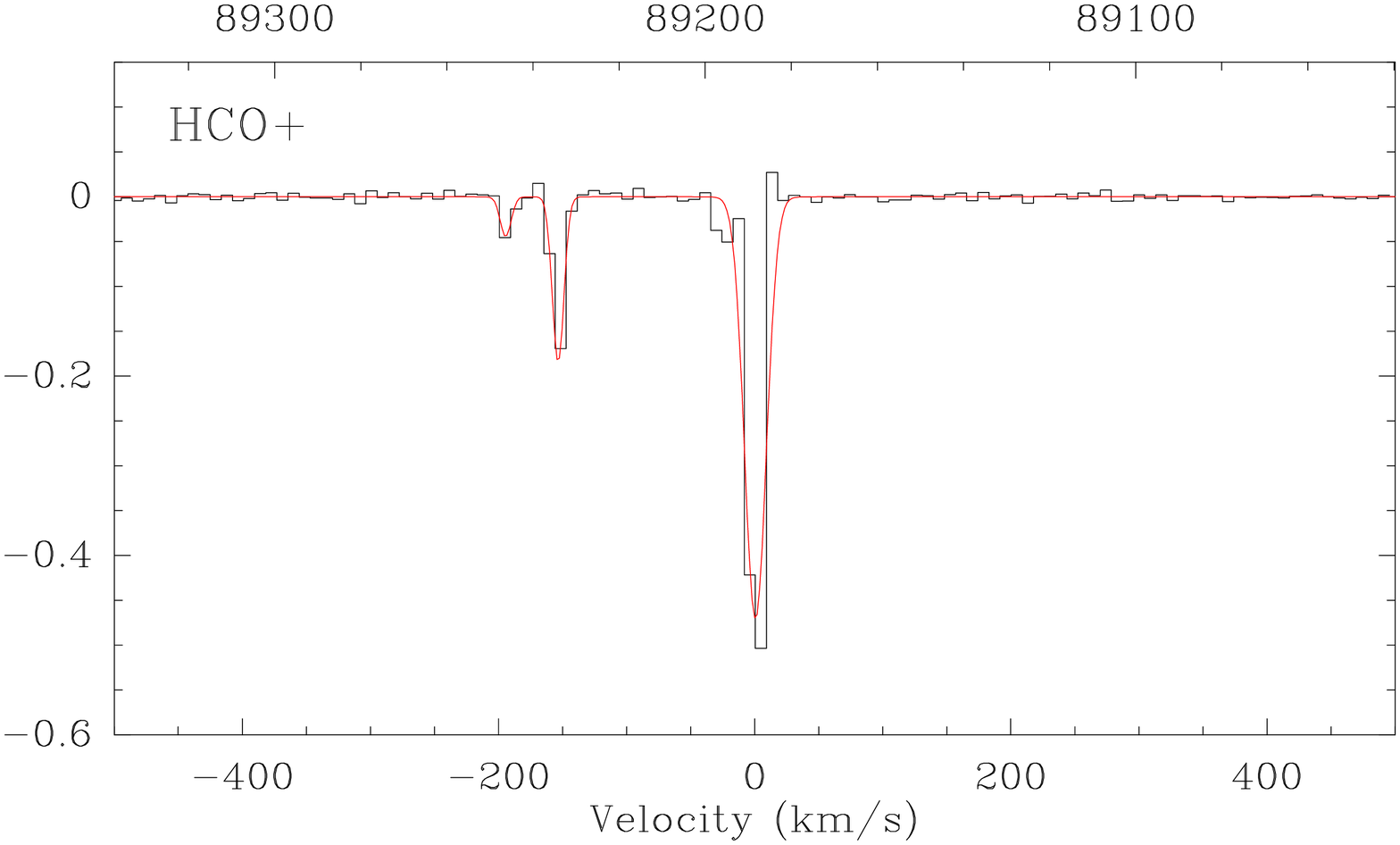}
}
}
\caption{The survey. In red: Gaussian fit for each detected species. For the species with HFS resolved (CCH, HCO, CN), the red line shows the velocity component at $\sim 0$ km s$^{-1}$ and the blue line shows the velocity component at $\sim -153$ km s$^{-1}$.}
\label{fit}
\end{figure*}

\begin{figure*}
\centering
\vbox{
\hbox{
\includegraphics[width=0.5\hsize]{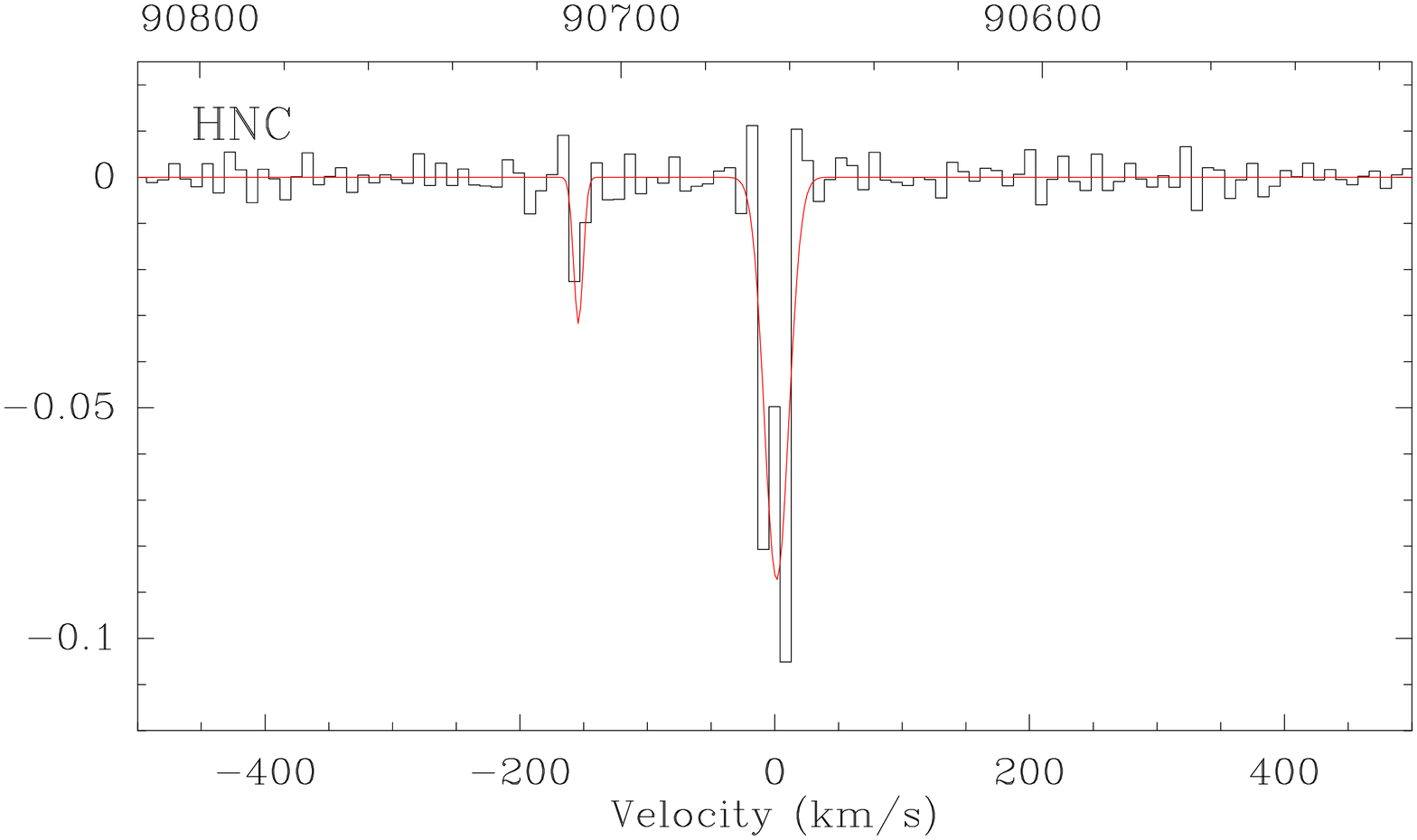}
\includegraphics[width=0.5\hsize]{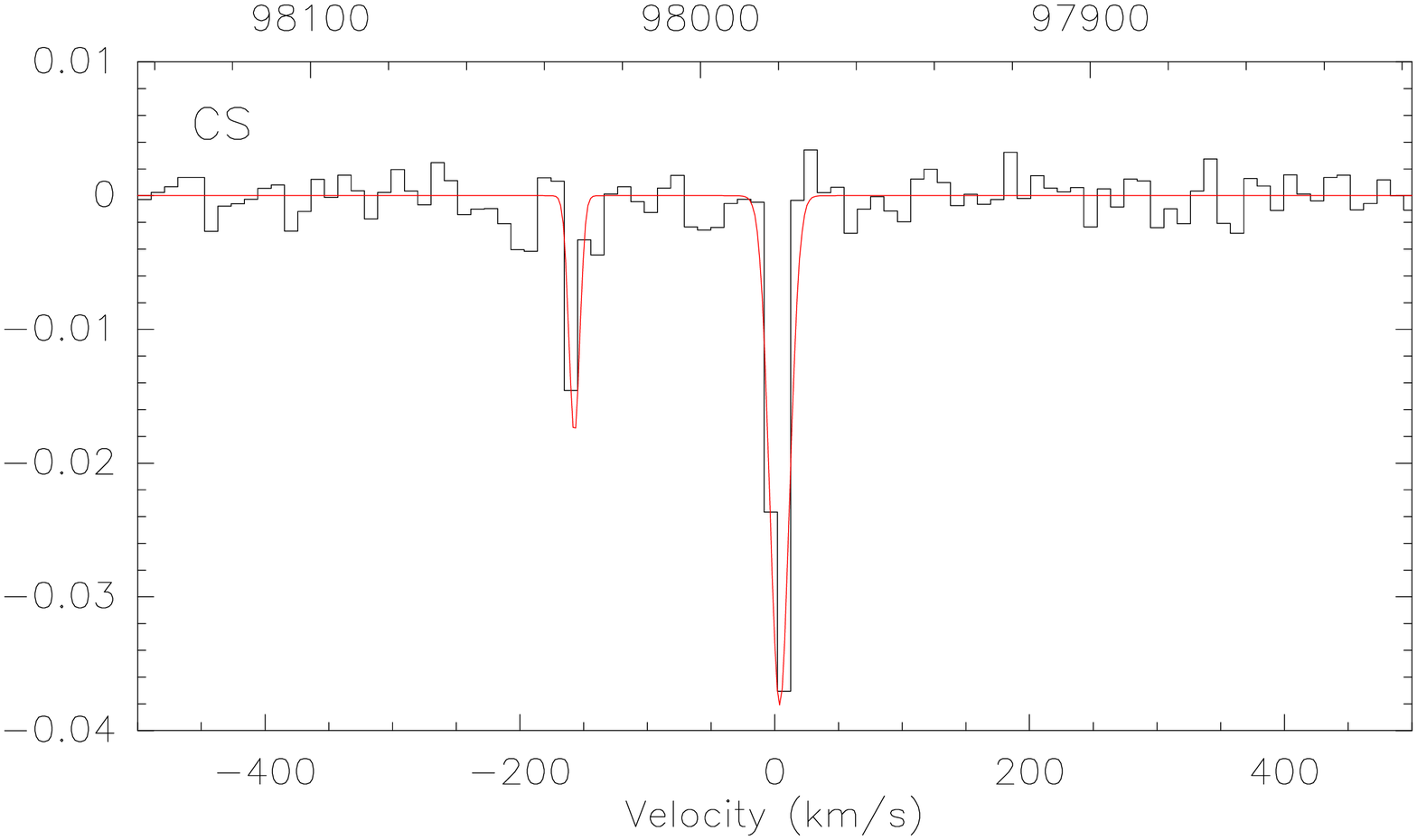}
}
\hbox{
\includegraphics[width=0.5\hsize]{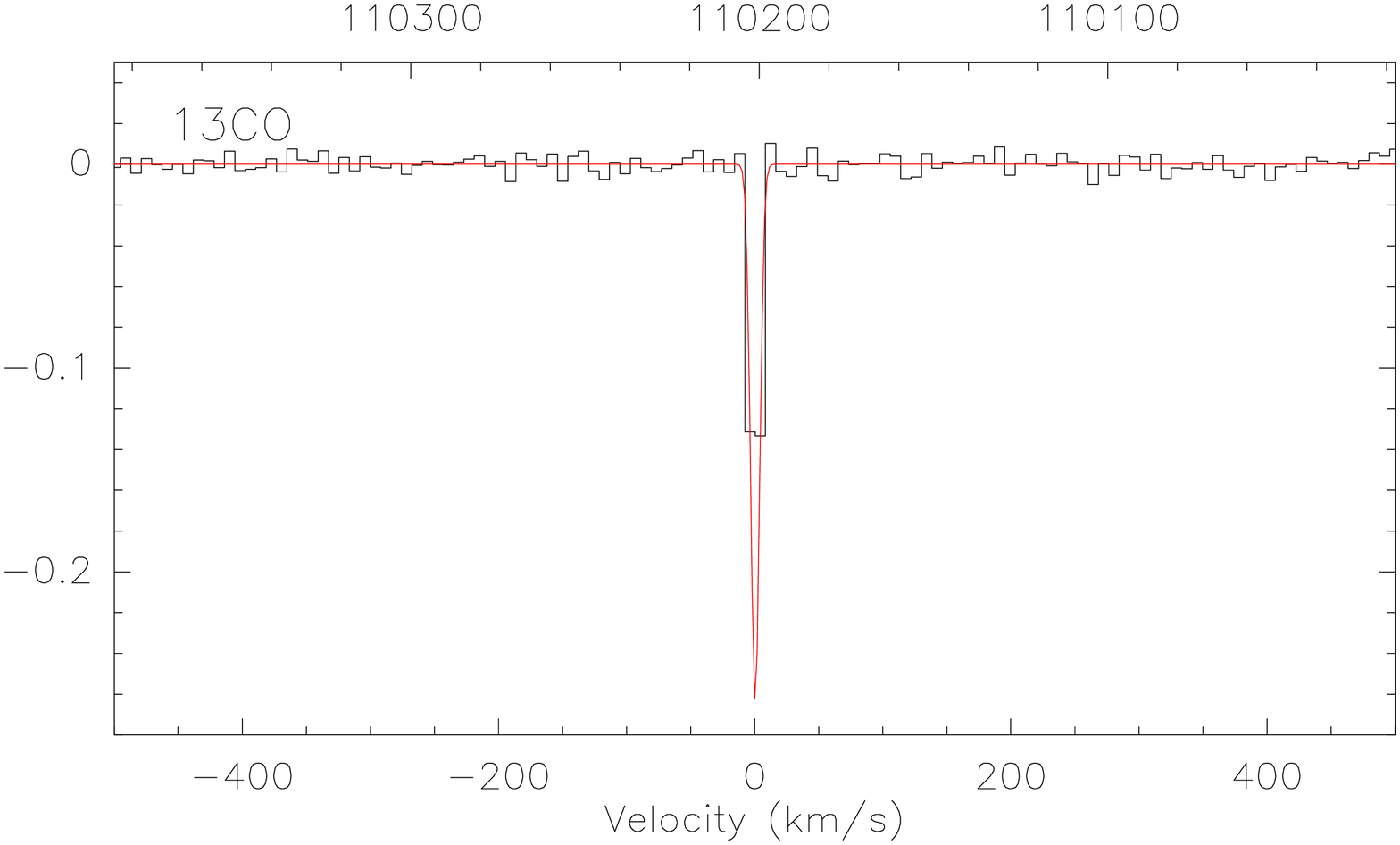}
\includegraphics[width=0.5\hsize]{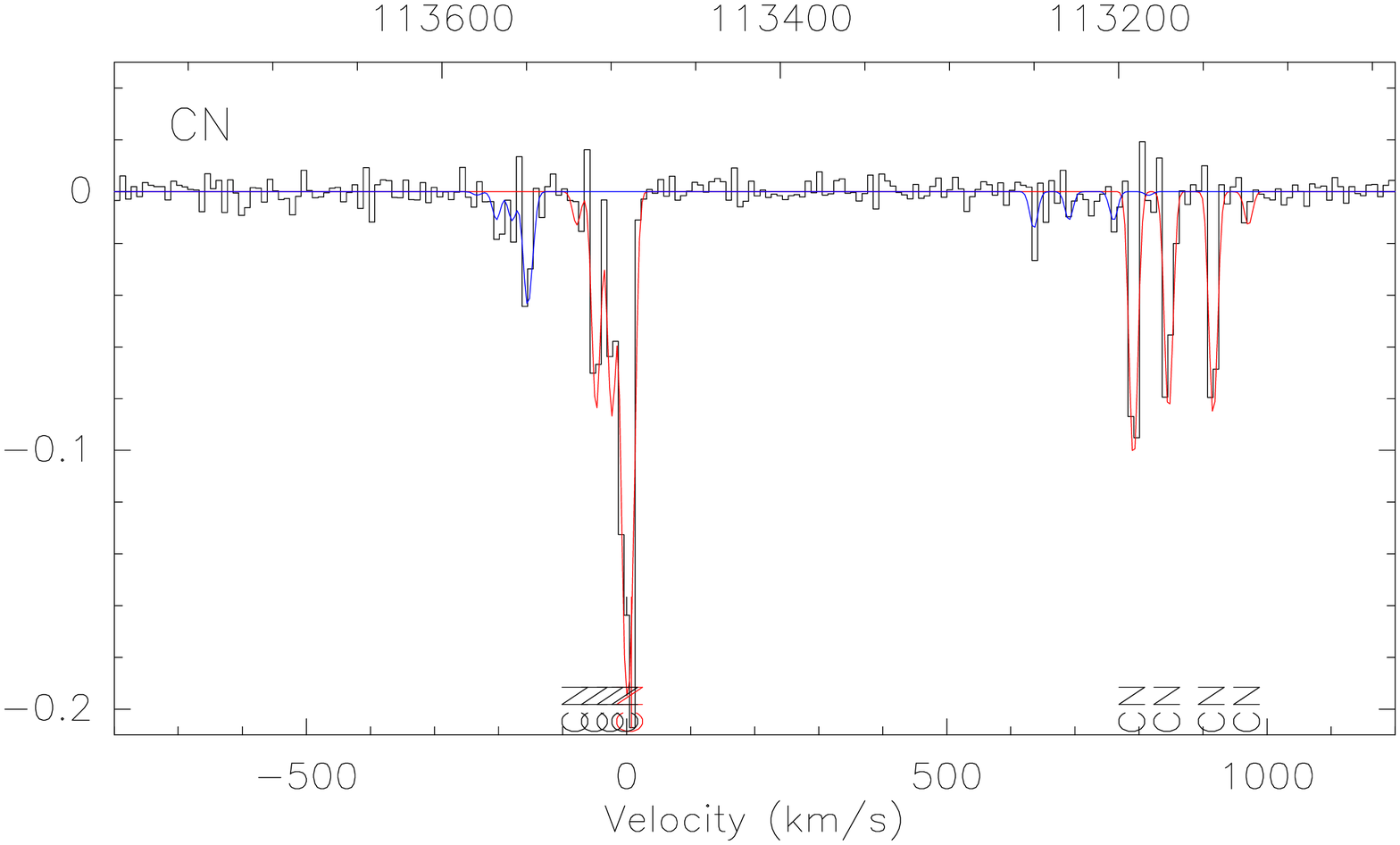}
}
}
\caption{The survey. In red: Gaussian fit for each detected species. For the species with HFS resolved (CCH, HCO, CN), the red line shows the velocity component at $\sim 0$ km s$^{-1}$ and the blue line shows the velocity component at $\sim -153$ km s$^{-1}$.}
\label{FigVibStab}
\end{figure*}
\end{appendix}

\end{document}